\documentclass[12pt]{article}
\usepackage[utf8]{inputenc}
\usepackage{geometry}
\usepackage{graphicx}
\usepackage{array}
\usepackage{subfig}
\usepackage{slashed}
\usepackage{amsmath}
\usepackage{amsfonts}
\usepackage{amssymb}
\usepackage{tensor}
\usepackage[cm]{fullpage}
\usepackage{cite}
\usepackage{epstopdf}
\usepackage{comment}
\usepackage{hyperref}
\usepackage{cancel}
\usepackage{mathrsfs}

\usepackage{cleveref}
\crefname{section}{§}{§§}
\Crefname{section}{§}{§§}

\usepackage{booktabs} 
\numberwithin{equation}{section}

\def\0{{(0)}}
\def\1{{(1)}}
\def\2{{(2)}}

\def\<{\langle }
\def\>{\rangle }

\def \ap{\alpha'}

\newcommand{\eq}{ \ \ = \ \ }

\newcommand{\bea}{\begin{eqnarray}}

\newcommand{\eea}{\end{eqnarray}}

\newcommand{\be}{\begin{equation}}
\newcommand{\ee}{\end{equation}}
\newcommand{\ba}{\begin{align}}
\newcommand{\ea}{\end{align}}

\newcommand{\W}{\mathcal{W}}

   \makeatletter
  \let\over=\@@over \let\overwithdelims=\@@overwithdelims
  \let\atop=\@@atop \let\atopwithdelims=\@@atopwithdelims
  \let\above=\@@above \let\abovewithdelims=\@@abovewithdelims
\renewcommand\section{\@startsection {section}{1}{\z@}%
                                   {-3.5ex \@plus -1ex \@minus -.2ex}%nn
                                   {2.3ex \@plus.2ex}%
                                   {\normalfont\large\bfseries}}

\renewcommand\subsection{\@startsection{subsection}{2}{\z@}%
                                     {-3.25ex\@plus -1ex \@minus -.2ex}%
                                     {1.5ex \@plus .2ex}%
                                     {\normalfont\bfseries}}

\newcommand{\beq}{\begin{equation}}
\newcommand{\eeq}{\end{equation}}
\newcommand{\beqa}{\begin{eqnarray}}
\newcommand{\eeqa}{\end{eqnarray}}
\newcommand{\beqar}{\begin{eqnarray*}}

\def\[{\[}
\def\]{\]}

\newcommand{\bd}[1]{\begin{fmffile}{#1}\begin{fmfgraph*}}
\newcommand{\ed}{\end{fmfgraph*}\end{fmffile}}

\begin{document}

\begin{titlepage}

\begin{flushright}
%\vskipcm
{\small  CERN--TH--2018--222},
{\small  MPP--2018--249},
{\small LMU--ASC 64/18}
\end{flushright}

\vspace{0.2cm}

\begin{center}

{\LARGE{\textsc{Bimetric, Conformal Supergravity and its}}}

\vskip0.3cm

{\LARGE{\textsc{Superstring Embedding}}}

%\vskip0.5cm

%{\LARGE{\textsc{Weyl Supergravity}}}

%\vskip0.4cm

% {\LARGE{\textsc{S-fold and Double Copy Construction}}}

\vspace{0.5cm}
{\large \bf Sergio Ferrara$^{a,b,c}$,   \, Alex Kehagias$^{d}$,\, Dieter L\"ust$^{e,f}$}
%\\
%{~}\\
%{\large  Dieter L\"ust$^{b,c}$,\ \ Diego Regalado$^b$  }

\vspace{0.5cm}

{\it

\vskip-0.3cm
\centerline{ $^{\textrm{a}}$ CERN, Theory Department,}
\centerline{1211 Geneva 23, Switzerland}
\medskip
\centerline{ $^{\textrm{b}}$ INFN, Laboratori Nazionali di Frascati,}
\centerline{Via Enrico Fermi 40, 00044 Frascati, Italy}
\medskip
\centerline{ $^{\textrm{c}}$ Department of Physics and Astronomy}
\centerline{and Mani L. Bhaumik Institute for Theoretical Physics, U.C.L.A}
\centerline{Los Angeles CA 90095-1547, U.S.A.}
\medskip
\centerline{ $^{\textrm{d}}$ Physics Division, National Technical University of Athens}
\centerline{ 15780 Zografou Campus, Athens, Greece}
\medskip
\centerline{ $^{\textrm{e}}$ Arnold--Sommerfeld--Center for Theoretical Physics,}
\centerline{Ludwig--Maximilians--Universit\"at, 80333 M\"unchen, Germany}
\medskip
\centerline{$^{\textrm{f}}$ Max--Planck--Institut f\"ur Physik,
Werner--Heisenberg--Institut,}
\centerline{ 80805 M\"unchen, Germany}

}

\end{center}

\vskip0.4cm
\abstract{{
We discuss the connection between Weyl$^2$ supergravity and superstrings and further discuss holography between 4-dimensional, ${\cal N}=4$ superconformal 
Weyl$^2$ supergravity and ${\cal N}=8$, higher spin-four theory on $AdS_5$.
The  Weyl$^2$ plus Einstein supergravity theory is a special kind of a bimetric gravity theory and consists of a massless graviton multiplet plus an additional massive spin-two supermultiplet.
Here, we argue that the additional
spin-two field and its superpartners originate from massive  excitations in the open string sector;  just like the ${\cal N}=4$ super Yang-Mills gauge fields, they
are localized on the world volume of D3-branes.
The ghost structure of the Weyl action should be considered as an artifact of the truncation of the infinitely many higher derivative terms underlying the massive spin 2 action.
In field theory, ${\cal N}=4$ Weyl$^2$ supergravity exhibits superconformal invariance in the limit of vanishing Planck mass. In string theory the
additional spin-two fields become massless in the tensionless limit. Therefore 
low string scale scenarios with large extra dimensions provide  (almost) superconformal field theories with almost massless open string spin-two fields. 
The full ${\cal N}=4$ scalar potential including the Yang-Mills matter multiplets is presented and the supersymmetric
vacua of Einstein Supergravity are shown, as expected, to be vacua of massive Weyl supergravity.
Other vacua are expected to exist
which are not vacua of Einstein supergravity.
Finally, we  identify certain spin-four operators on the 4-dimensional boundary theory that could be 
the holographic duals of spin-four fields in the bulk.
}}

\end{titlepage}

\newpage

\tableofcontents
\break

\section{Introduction}

It is well known that the effective action of string theory is given in terms Einstein gravity,  coupled to matter fields plus in finite series
of higher derivative terms, which in particular contain an infinite series of higher curvature terms, which are suppressed by appropriate powers of the string scale $M_s=(\alpha')^{-1}$.
In the so-called field theory limit of sending $\alpha'\rightarrow 0$, all higher string modes decouple and all higher derivative interactions disappear, and the effective theory is just given
by the Einstein-Yang-Mills-theory. Particular string examples of those theories are brane-world models, where the Yang-Mills degrees of freedom are localized on the
world-volumes of stack of D-branes, and where the gravitational fields, namely the metric field $g_{\mu\nu}$ and its partners, correspond to closed strings, which propagate within the entire
ten-dimensional bulk space. Here will will consider the simplest case, namely a stack of N D3-branes, i.e. the open string Yang-Mills sector is confined
on the 4D world-volume of the D3-branes.

Now, when considering also higher curvature terms up to four derivatives \cite{Stelle,Boulware,David,Horowitz,Deser,tHooft,Maldacena,LPS}, it is again well known that the $R^2$ action and the so-called 
Weyl$^2$ action propagate additional degrees of freedom:
for $R^2$ there is an additional scalar mode and for Weyl$^2$ there exist an additional spin two field, denoted by $w_{\mu\nu}$. In this paper we will discuss the physics connected
to the Weyl$^2$ action and to spin-two field $w_{\mu\nu}$ and in particular the question how do they arise in string theory. Since the theory contains two spin-two metric fields, 
namely $g_{\mu\nu}$ and $w_{\mu\nu}$, it is a particular example of a bimetric gravity theory       \cite{Hassan:2011tf}.
As we will discuss the second spin-two mode $w_{\mu\nu}$ is not contained in standard closed string gravitational sector, 
but it corresponds to the first massive open string excitations, namely to the massive excitations of the open string Yang-Mills gauge fields.
Therefore these massive fields $w_{\mu\nu}$ are also localized on the world volume of D3-branes,
 and the effective Weyl$^2$ is an entirely four-dimensional action on the world-volume of the D3-branes.
 As we will discuss, performing a particular  scaling  limit, the closed string gravitational modes decouple, and one is left with an
 effective 4D theory of only open string modes, namely massless ${\cal N}=4$ super Yang-Mills gauge theory plus (almost) massless ${\cal N}=4$ super-Weyl$^2$ theory, whose
 spectrum was recently constructed in \cite{Ferrara:2018wqd}. Hence in this limit the theory becomes (almost) superconformal invariant.
 Note that superconformal Weyl$^2$ gravity \cite{FK1,K2,FTs3,FvP},  only exists for numbers of supersymmetries ${\cal N}\leq 4$, just like superconformal Yang-Mills gauge theories also only exist for ${\cal N}\leq 4$ \cite{deWit:1978pd}.
 This fact confirms our observation that Weyl$^2$ gravity is not originating entirely from closed strings, but is
 an effective open string theory, localized on D3-branes.

These theories are also of phenomenological interest, namely in the context of the  low string scale scenario together with large extra dimensions, which allows
for unique predictions for the production of the massive open string excitations at particle physics collider machines \cite{Lust:2008qc}.
Namely, following the discussion of this paper, the low string scale scenario with 
light, open string spin-two excitations is a (almost) superconformally invariant field theory.

As we will argue in the last part of the paper, the 4D (almost) super-conformal invariant Weyl supergravity theory allows for an holographic description  in terms of closed string modes in an $AdS_5$ 
bulk theory. In contrast to the standard AdS/CFT  correspondence between massless open string Yang-Mills gauge theory in the 4D boundary and  supergravity in the 5D bulk,
the holographic description of the (almost) massless spin-two fields on the boundary is given by (almost) massless spin-four fields in the higher-dimensional bulk.

The structure of the paper is as follows: In section 2 we describe  Weyl supergravity coupled to super Yang-Mills theory. In section 3, we present a string theory realization of the theory and in section 4 we present some of its holographic aspects. Finally, section 5 contains our conclusions.   

%%%%%%%%%%%%%

\section{Field theory: (Super)-Yang-Mills plus (Super)-Weyl gravity}

\subsection{Bosonic case}

The most general formulation of  Einstein plus curvature-square  gravity is described  by an action  containing the standard Einstein term plus the
following two terms being second order in the curvature tensor\footnote{There are two more linear combination of quadratic curvature terms, namely the Gauss-Bonnet and the  Hirzebruch–Pontryagin action.
However in four-dimensions these are total derivatives and hence we neglect them in the following. Similar considerations exist also in the supersymmetric case \cite{CFS,FFKL}.}:
\begin{eqnarray}
S=\int_{\cal M} d^4 x\sqrt{-g}\Big{(}M_P^2 R+c_1 W_{\mu\nu\rho\sigma}W^{\mu\nu\rho\sigma}+c_2 R^2\Big{)}. \label{W}
\end{eqnarray}
More details can be e.g. found in \cite{Alvarez-Gaume:2015rwa,Salvio:2018crh}.
The first term with $W_{\mu\nu\rho\sigma}$ being the Weyl tensor
\begin{eqnarray}
W_{\mu\nu\rho\sigma}=R_{\mu\nu\rho\sigma}+
g_{\mu[\sigma}R_{\rho]\nu}+g_{\nu[\rho}R_{\sigma]\mu}+\frac{R}{3} g_{\mu[\rho}g_{\sigma]\nu}
\end{eqnarray}
is conformally invariant, whereas the $R^2$ term 
is
 only scale invariant.  
Indeed,  
 the conformal transformation
\begin{eqnarray}
g_{\mu\nu}\to \widehat{g}_{\mu\nu}=\Omega^2 g_{\mu\nu}, \label{conf}
\end{eqnarray}
leaves the Weyl tensor inert
\begin{eqnarray}
\widehat{W}^\mu_{~\nu\rho\sigma}= W^\mu_{~\nu\rho\sigma} ,
\end{eqnarray}
whereas the curvature scalar transforms as 
\begin{eqnarray}
\widehat{R}= \Omega^{-2}R - 6 \Omega^{-3} g^{\mu\nu}\nabla_\mu 
\nabla_\nu \Omega.  
\end{eqnarray} 
The two couplings $c_i$ in (\ref{W}) are dimensionless.
As discussed in \cite{Alvarez-Gaume:2015rwa}, the $R^2$ action only propagates a scalar mode in flat four-dimensional space-time $\mathrm{R} ^{1,3}$. 
%On the other, as we will further discuss,
%the Einstein plus $W^2$ action propagates the standard spin-two graviton $g_{\mu\nu}$ plus an additional massive spin-two field $w_{\mu\nu}$.
Since we are in particular interested in spin-two fields and not to the  additional scalar mode in the string spectrum,  the $R^2$ action is not relevant for us, and we will
set the coupling $c_2=0$. 
However, the action (\ref{W}) with $c_2=0$ is not conformal invariant since the Einstein-term is not invariant under conformal transformations. 
Therefore the Einstein-term can be regarded as the mass term in this theory, i.e. a mass deformation, which explicitly breaks conformal invariance.

%Therefore we are dealing with a theory with two spin-two fields, one massless and one massive, which is an particular example of a bimetric gravity theory
%\cite{Gording:2018not}.
The propagator  of the Einstein-Weyl$^2$ theory \cite{Stelle} described by 
\begin{eqnarray}
S=\int_{\cal M} d^4 x\sqrt{-g}\Big{(}M_P^2 R+\frac{1}{2g_W^2} W_{\mu\nu\rho\sigma}W^{\mu\nu\rho\sigma}\Big{)}, \label{EW}
\end{eqnarray}
 is given by the following expression 
\begin{equation}
\Delta_{\mu\nu\rho\sigma}=\Delta(k)  P_{\mu\nu\rho\sigma},%={g_W^2\over k^2(k^2-g_W^2M_P^2)} P_{\mu\nu\rho\sigma},
\label{propagator}
\end{equation}
where 
\begin{eqnarray}
\Delta(k)={g_W^2\over k^2(k^2-g_W^2M_P^2)}, \label{d}
\end{eqnarray}
and 
\begin{eqnarray}
P_{\mu\nu\rho\sigma}=\frac{1}{2}\Big(\theta_{\mu\rho}\theta_{\nu\sigma}+\theta_{\mu\sigma}\theta_{\nu\rho}\Big)-\frac{1}{3}
\theta_{\mu\nu}\theta_{\rho\sigma}\, , ~~~
\end{eqnarray}
with
\begin{eqnarray}
\theta_{\mu\nu}=\eta_{\mu\nu}-\frac{k_\mu k_\nu}{k^2}
\end{eqnarray}
 the usual transverse  vector projection operator. Note that the propagator (\ref{propagator}) for $M_P^2=0$ (i.e., pure Weyl\textsuperscript{2} theory) exhibits the  conformal $1/k^4$ behaviour. When the Einstein terms is present, we  can equivalently  write $\Delta(k)$ as
 %(\ref{propagator}) as 
 \begin{eqnarray}
%  \Delta_{\mu\nu\rho\sigma}=\Delta(k)  P_{\mu\nu\rho\sigma}=\left(-\frac{1}{M_P^2}\frac{1}{k^2}+
%  \frac{1}{M_P^2}\frac{1}{k^2-g_W^2 M_P^2}\right) P_{\mu\nu\rho\sigma}, \label{pp}
\Delta(k)=-\frac{1}{M_P^2}\frac{1}{k^2}+
 \frac{1}{M_P^2}\frac{1}{k^2-g_W^2 M_P^2},\label{pp}
  \end{eqnarray}
  where the massless helicity-$\pm 2$ graviton is easily identified in the first term of (\ref{pp}). Moreover, we see that there is also a massive spin-2 state (the second term in (\ref{pp})) with mass given by the pole at $k^2=g_W^2 M_P^2$ which however has opposite residue to the usual massles graviton, and therefore describes a ghost spin-2 state. This  shows that the theory contains as propagating degrees the standard, massless spin-two graviton $g_{\mu\nu}$ plus an additional massive spin-two field $w_{\mu\nu}$.

\vskip0.2cm
Actually, an alternative way to see this is to write down a particular bimetric gravity theory with two spin-two fields $g_{\mu\nu}$ and  $w_{\mu\nu}$ with the following
two-derivative action \cite{Bergshoeff:2009hq}:
\begin{eqnarray}
S=\int_{\cal M} d^4 x\sqrt{-g}\Big{(}M_P^2 R(g)+{2M_P}G_{\mu\nu}(g)w^{\mu\nu}-M_W^2(w^{\mu\nu}w_{\mu\nu}-aw^2)\Big{)}. \label{bimetric}
\end{eqnarray}
Here $G_{\mu\nu}=R_{\mu\nu}-1/2 R\,g_{\mu\nu}$ is the Einstein-tensor constructed from the metric $g_{\mu\nu}$ and the last term is a mass term for the second metric $w_{\mu\nu}$. In general the action 
propagates also a massive scalar mode. However setting the parameter $a=1$, the scalar mode disappears and the action contains a massless spin-two field $g_{\mu\nu}$ 
plus a massive spin-two field $w_{\mu\nu}$. Note that the two-derivative kinetic term for $w_{\mu\nu}$ is hidden in the coupling $G_{\mu\nu}(g)w^{\mu\nu}$, which can be seen by performing two partial integrations on this term. However after the partial integrations the kinetic term for $w_{\mu\nu}$ has the wrong sign,
i.e. $w_{\mu\nu}$ is a ghost-like field. Now using the equation of motion 
\begin{equation}
{\delta S\over \delta w_{\mu\nu}}\quad\Rightarrow\quad w_{\mu\nu}=\frac{M_P}{M_W^2}\left(R_{\mu\nu}(g)-{1\over 6}g_{\mu\nu}R\right)\, ,\label{eqmotion}
\end{equation}
and plugging the solution for $w_{\mu\nu}$ back into the action (\ref{bimetric}), one can show \cite{Gording:2018not} that the resulting action is (classically) equivalent to the four-derivative $W^2$ action
in eq.(\ref{EW}) by using the fact that 
\begin{eqnarray}
W_{\mu\nu\rho\sigma}W^{\mu\nu\rho\sigma}=GB+2(R_{\mu\nu}R^{\mu\nu}-\frac{1}{3}R^2), \label{GBW}
\end{eqnarray}
where $GB=R_{\mu\nu\rho\sigma}R^{\mu\nu\rho\sigma}-4R_{\mu\nu}R^{\mu\nu}+R^2$ is the Gauss-Bonnet term. 
The bimetric gravity action (\ref{bimetric}) for $w_{\mu\nu}$ can be made ghost-free by adding an infinite number of terms with a finite number of parameters to it.
As shown \cite{Gording:2018not}, this procedure  is equivalent to adding to the $W^2$ action an infinite number of higher derivative terms, which resemble to additional parameters
of the ghost-free bimetric gravity theory. In other words,  the ghost nature of the massive spin-2 excitation is an artifact of the higher derivative truncation to fourth order.
En passant, let us mention theat for $a\neq 1$, the action (\ref{bimetric}) is (classically) equivalent to the action (\ref{W}) with 
\begin{eqnarray}
c_1=\frac{1}{2g_W^2},~~~c_2=\frac{a-1}{4a-1}\frac{1}{3g_W^2} . 
\end{eqnarray}
Therefore, only for $a=1$ the scalar mode associated to the $R^2$ term is absent. 

\subsection{Supersymmetric case}

The above method can also be implemented in a supersymmetric setup \cite{CFS}. For this, we need to recall that the graviton $h_{\mu\nu}$ sits in a real vector superfield $\Phi_\mu$ with expansion (in Wess-Zumino gauge)
\begin{eqnarray}
\Phi_\mu=\overline\theta \sigma^\nu \theta (h_{\mu\nu}+A_{\mu\nu})+\frac{1}{2}\overline \theta ^2 \theta^2 A_\mu+\cdots,
\end{eqnarray}
where $A_{\mu\nu}$ and $A_\mu$ are  the antisymmetric two-form and one-form  fields of new-minimal supergravity, respectively. We can then define the real linear superfield 
$E_\mu$ as 
\begin{eqnarray}
E^{\mu}=\frac{1}{2} \epsilon^{\mu\nu\rho\sigma} \overline D\sigma_\nu  D \partial_\rho\Phi_\sigma, 
\end{eqnarray}
which contains the Einstein term  
\begin{eqnarray}
E^\mu=\overline\theta \sigma_\nu \theta (G^{\mu\nu}+\partial_\lambda F^{\lambda\nu\mu}+\frac{1}{2}\epsilon^{\nu\mu\rho\sigma}F_{\rho\sigma})+\cdots,
\end{eqnarray}
with $F_{\mu\nu\rho}=\partial_\mu A_{\nu\rho}+\partial_\rho A_{\mu\nu}+\partial_\nu A_{\rho\mu}$ and $F_{\mu\nu}=\partial_\mu A_\nu-\partial_\nu A_\mu$ the field strengths of the auxiliaries $A_{\mu\nu}$ and $A_\mu$, respectively. We need also to define the
Riemann multiplet $R_{\mu\nu}$ with components expansion
\begin{eqnarray}
R_{\mu\nu}=\frac{1}{2}\psi_{\mu\nu}+\frac{i}{2}\theta^2 \sigma^\rho
\partial_\rho\overline \psi_{\mu\nu}-\frac{i}{2}\theta F_{\mu\nu}
-\frac{i}{4} \sigma^{\kappa\lambda}\theta 
(R_{\kappa\lambda\mu\nu}+\partial_\nu F_{\mu\kappa\lambda}-\partial_\mu F_{\nu\kappa\lambda}).
\end{eqnarray}
The Weyl tensor $\W_{\mu\nu\rho\sigma}$ is contained then in the Weyl multiplet $W_{\mu\nu}$ defined as 
\begin{eqnarray}
W_{\mu\nu}=\frac{1}{8}\left(\sigma^{\kappa\lambda}\sigma_{\mu\nu}+\frac{1}{3}\sigma_{\mu\nu}\sigma^{\kappa\lambda}\right)R_{\kappa\lambda}, 
\end{eqnarray}
as can be seen from its components expansion
\begin{eqnarray}
W_{\mu\nu}= \frac{1}{16}\left(\sigma^{\kappa\lambda}\sigma_{\mu\nu}+\frac{1}{3}\sigma_{\mu\nu}\sigma^{\kappa\lambda}\right)\psi_{\kappa\lambda}-i\sigma^{\kappa\lambda}\theta W_{\kappa\lambda\mu\nu}+\cdots\, .
\end{eqnarray}
In terms of the real vector superfield $\Phi_\mu$, the Riemann and Weyl multiplets can be written (in spinor notation), 
\begin{eqnarray}
R_{\mu\nu\alpha}=-\frac{1}{8}\overline D ^2D_\alpha \left(\partial_\mu\Phi_\nu-\partial_\nu\Phi_\mu\right), ~~~W_{\alpha\beta\gamma}=
\frac{1}{16}\overline D ^2 D_{(\alpha}\partial_\beta^{\dot{\alpha}}
\Phi_{\gamma)\dot{\alpha}}. 
\end{eqnarray}
The action (\ref{W}) (with $c_2=0$) is contained in the bosonic part of the supersymmetric Lagrangian (with $M_P=1$ here)
\begin{eqnarray}
\mathcal{L}=\int d^4 \theta \Phi_\mu E^\mu +8 c_1 {\rm Re}\int d^2 \theta W_{\mu\nu}W^{\mu\nu} .
\end{eqnarray}
The first term contains the Einstein term and the second the Weyl\textsuperscript{2}. 
A supersymmetric generalization of (\ref{GBW}) exists and it is written as
\begin{eqnarray}
W_{\mu\nu}W^{\mu\nu}=SGB-\frac{1}{8}\overline D ^2 (E_\mu E^\mu)+\frac{1}{3} W^2, \label{ww}
\end{eqnarray}
where 
\begin{eqnarray}
W=\frac{1}{2}\sigma^\mu \overline D E_\mu
\end{eqnarray}
and $SBG$ is the supersymmetric counterpart of the usual Gauss-Bonnet term and it is such that in the real and imaginary parts of its highest $\theta^2$ component are the 
Hirzebruch-Pontryagin and Gauss-Bonnet terms, respectively. We may then write (\ref{ww}) as 
 \begin{eqnarray}
 \mathcal{L}=\int d^4 \theta \left(\Phi_\mu E^\mu -4c_1 E_\mu E^\mu\right)+\frac{8}{3} c_1 {\rm Re}\int d^2 \theta W^2. \label{2}
 \end{eqnarray}
We may linearize in $E_\mu$ and $W$ the above  Lagrangian by introducing a real vector superfield $V_\mu$ and a superfield $H$ so that 
 \begin{eqnarray}
 \mathcal{L}=\int d^4 \theta \left(\Phi_\mu E^\mu +2V_\mu E^\mu
 +\frac{1}{4c_1} V_\mu V^\mu\right)- {\rm Re}\int d^2\theta \left(2W \overline D ^2 H+\frac{3}{8c_1}(\overline D ^2 H)^2\right). \label{3}
 \end{eqnarray}
 Then, after performing first the shift $\Phi_\mu\to \Phi_\mu-V_\mu$ and after the conformal transformation $\Phi_\mu\to \Phi_\mu+
 \overline D \sigma_\mu H+D\sigma_\mu \overline H$, we get that 
  the supersymmetric action (\ref{2}) is classically equivalent to  
 \begin{eqnarray}
 \mathcal{L}&=&\int d^4 \theta\Phi_\mu E^\mu -\int d^2 \theta
 \left(V_\mu E^\mu
 -\frac{1}{4c_1} V_\mu V^\mu+\frac{3}{8} L^2\right)\nonumber \\
&& - {\rm Re}\int d^2\theta \left(2W \overline D ^2 H+\frac{3}{8c_1}(\overline D ^2 H)^2\right), \label{4}
 \end{eqnarray}
where $L=D\overline{D}^2 H-\overline D D^2 \overline H$ \cite{CFS}. 
From the above Lagrangian we see that the first term in the first line describes a physical massless $(2,\frac{3}{2})$ graviton multiplet $(\Phi_\mu)$, whereas the second term in the first line describes a  massive $(2,\frac{3}{2},\frac{3}{2},1)$  multiplet ($V_\mu$)  with mass square
$m^2=1/4c_1$ \cite{FZ}. The latter multiplet is not physical as its Lagrangian term opposite sign from the massless multiplet and therefore it is a ghost massive spin-2 multiplet.

\subsection{Including Gauge Fields}
%\vskip0.2cm
Now, we will also include 
 a four-dimensional bosonic Yang-Mills $U(N)$ gauge theory, which is coupled to Einstein gravity. Then the action up to four orders in derivatives  has the following form:
\begin{eqnarray}
S
=\int d^4 x\sqrt{-g}\Big{(}-{1\over 4 g_{YM}^2}  F_{\mu\nu}^aF^{a\,\mu\nu}+{1\over 2 g_W^2}  W_{\mu\nu\rho\sigma}W^{\mu\nu\rho\sigma}+M_P^2 R\Big{)}
. \label{W4}
\end{eqnarray}
$ F_{\mu\nu}^a$ is the standard Yang-Mills field strength
and  $g^2_W$ and   $g^2_{YM} $ are dimensionless couplings.
 The Yang-Mills term and the Weyl$^2$-term in the action possess (classical) conformal invariance,  whereas again the  Einstein-term can be regarded as the mass term in this theory.

Let us recall   the propagating modes corresponding to this action. Specifically, there are three kinds of propagating modes \cite{Stelle,FZ,FGvN,Rieg}:

\vskip0.5cm
\noindent

\vskip0.2cm
\noindent
(i) A  massless helicity-$\pm2$ graviton $g_{\mu\nu}$. This is the standard massless spin-two graviton.

\vskip0.2cm
\noindent
(ii) Massless  $U(N)$  gauge bosons $A_\mu^a$.

\vskip0.2cm
\noindent
(iii)  A massive spin-two particle $w_{\mu\nu}$ with mass  
\begin{equation}
M_W=g_WM_P\, .\label{mw}
\end{equation}
It is related to the Weyl$^2$ term in the action. In fact as mentioned, this massive spin two particle is a ghost, destroying
unitarity, but we will neglect this problem in the following and we will comment on it
only in the conclusions.  We will call this part of the spectrum the non-standard sector of the theory.
\vskip0.2cm
\noindent
The Einstein plus (Weyl)$^2$ gravity theory contains seven propagating degrees of freedom.
As already explained, this part of the theory can be considered as a bimetric theory of gravity with 
 two spin-two fields, namely one the standard massless graviton $g_{\mu\nu}$ plus the non-standard massive spin-two field $w_{\mu\nu}$.
 As we will discuss in the following, in string theory the graviton $g_{\mu\nu}$ originates from the closed string sector and lives in the bulk space, whereas the 
 spin-two field $w_{\mu\nu}$ as well as the Yang-Mills gauge bosons $A_{\mu}^a$ come from the open string sector and will be localized on the world-volume of a stack
 of D3-branes.

%{\bf Remark on the energy momentum tensor.}

\vskip0.2cm
\noindent
In the following we will consider the following three limits.
% in the space of the three couplings $M_P, g_{YM}, g_W$.
Later we will see how these limits are realized in string theory.

%\vskip0.2cm
%\noindent{\sl (A) Einstein/Yang-Mills limit}
%\vskip0.2cm

%\noindent
%\subsection{Massless limit}
%First we consider  the  limit: 
%\begin{equation}
% M_P={\rm fixed}\, , \quad M_W,g_W\rightarrow\infty\, .
%\end{equation}
%In this limit the second spin-two field $w_{\mu\nu}$ will become infinitely heavy and decouples from the theory.
%The propagator $\Delta(k)$ in eq.(\ref{propagator}) becomes in this limit
%\begin{equation}
%\Delta(k)\rightarrow - {1\over k^2M_P^2}\, .
%\end{equation}
%Therefore   we recover standard Einstein plus Yang-Mills gauge theory:
%\begin{eqnarray}
%S= \int d^4 x\sqrt{-g} \Big{(}M_P^2 R-{1\over 2 g_{YM}^2}  F_{\mu\nu}^aF^{a\,\mu\nu}\Big{)}\, . \label{EYM}
%\end{eqnarray}

 \vskip0.2cm
\noindent{\sl (A) Decoupling of gravity, i.e. Yang-Mills limit}
\vskip0.2cm

\noindent
%\subsection{Complete decoupling of gravity, i.e. Yang-Mills limit}
First we consider  the infinite mass limit
\begin{equation}
 M_P\rightarrow \infty
 %\, , \quad g_W={\rm fixed}
 \,.
\end{equation}
In this limit gravity becomes non-dynamical and decouples from the theory. In fact, 
for non-zero coupling $g_W$,  both spin-two particles completely decouple, since the spin-two particle $w_{\mu\nu}$ becomes infinitely heavy.
Alternatively one can keep $M_W$ finite, which implies that $g_W\rightarrow 0$, i.e. the spin-two Weyl modes are very weakly coupled.

\vskip0.2cm
\noindent{\sl (B) Massless bigravity limit}
\vskip0.2cm

\noindent
%\subsection{Massless limit}
Second we consider  the massless limit, namely the limit of vanishing Planck mass\footnote{The massless limit was also discussed in the context of bimetric theories in
\cite{Gording:2018not,Hassan:2012gz}.}: 
\begin{equation}
 M_P\rightarrow 0
 %\, , \quad g_W={\rm fixed}
 \, .
\end{equation}
The propagator $\Delta(k)$ now becomes 
\begin{equation}
\Delta(k)\rightarrow {g_W^2\over k^4}\, . \label{d1}
\end{equation}
In this limit the second spin-two field $w_{\mu\nu}$ will become massless and we deal with massless Weyl gravity. Therefore, for finite $M_P$ there is a Higgs effect with respect to $w_{\mu\nu}$, and in the massless limit
the degrees of freedom of $w_{\mu\nu}$ will arrange themselves into proper massless fields (see below).
In this limit  we deal with Yang-Mills gauge theory  plus   Weyl$^2$ theory with  action
\begin{eqnarray}
S= \int d^4 x\sqrt{-g} \Big{(}-{1\over 4 g_{YM}^2}  F_{\mu\nu}^aF^{a\,\mu\nu}+{1\over 2 g_W^2}W_{\mu\nu\rho\sigma}W^{\mu\nu\rho\sigma}\Big{)}\, . \label{W2}
\end{eqnarray}

\vskip0.5cm\noindent
This theory possesses conformal invariance and it propagates the following degrees of freedom:
 \vskip0.3cm
\noindent
(i) The standard  massless, closed string spin-two graviton $g_{\mu\nu}$, corresponding to a planar wave in Einstein gravity.

\vskip0.2cm
\noindent
(ii) Massless open string $U(N)$  gauge bosons $A_\mu^a$.

\vskip0.2cm
\noindent
(iii)  In the non-standard sector there is massless open string spin-two ghost particle $w_{\mu\nu}$, which corresponds to a non-planar wave.  In addition there is a massless open string vector $w_\mu$, which originates from the 
$\pm 1$ helicities of the massive $w_{\mu\nu}$ particle. However note that the helicity zero component of $w_{\mu\nu}$
%the so-called  dilaton,  
does not correspond to a physical, propagating mode in the
massless limit, since it can be gauged away by the conformal transformations (\ref{conf}).

% \vskip0.2cm
%\noindent{\sl (C) Decoupling of gravity, i.e. Yang-Mills limit}
%\vskip0.2cm

%\noindent
%\subsection{Complete decoupling of gravity, i.e. Yang-Mills limit}
%Now we consider  the infinite mass limit
%\begin{equation}
% M_P\rightarrow \infty\, , \quad g_W={\rm fixed}\,.
%\end{equation}
%In this limit gravity becomes non-dynamical and decouples from the theory. In fact, 
%for non-zero coupling $g_W$,  both spin-two particles completey decouple, since the spin-two particle $w_{\mu\nu}$ becomes infinitely heavy.
%Alternatively one can keep $M_W$ finite, which implies that $g_W\rightarrow 0$, i.e. the spin-two Weyl modes are very weakly coupled.
%So one is left with a pure 
% Yang-Mills plus   gauge theory:
%\begin{eqnarray}
%S= \int d^4 x\sqrt{-g} {1\over 2 g_{YM}^2}  F_{\mu\nu}^aF^{a\,\mu\nu}\, . 
%\end{eqnarray}

%\subsection{Einstein plus Yang-Mills limit}
%Next we consider  the infinite coupling limit
%\begin{equation}
%(C):\qquad g_W\rightarrow \infty\, .
%\end{equation}
%Keeping $M_P$ finite only the massive spin-two particles  decouple, since they become infinitely heavy.
%So one is left with standard Einstein plus Yang-Mills gauge theory:
%\begin{eqnarray}
%S
%=\int d^4 x\sqrt{-g}\Big{(}{1\over 2 g_{YM}^2}  F_{\mu\nu}^aF^{a\,\mu\nu}+M_P^2 R\Big{)}
%. \label{W4}
%\end{eqnarray}

 \vskip0.2cm
\noindent{\sl (C) Light spin-two plus massless Yang-Mills limit}
\vskip0.2cm

 %Gravitational Higgs effect

%\subsection{Massive spin-two plus massless Yang-Mills limit}
Now we consider  the double scaling limit 
\begin{equation}
 M_P\rightarrow \infty\quad{\rm and}\quad g_W\rightarrow 0\, \quad{\rm with}\quad M_W<<M_P\, .
\end{equation}
Therefore the coupling $g_W$ must vanish  faster than $M_P^{-1}$.
 In this limit the massless graviton decouples from theory, i.e. the standard gravitational 
sector gets decoupled from the massless non-standard spin-two sector. So one is left which an action that contains the massless Yang-Mills gauge fields $A_\mu$ as well as the 
(almost) massless spin-two fields $w_{\mu\nu}$.  The propagator has the leading behaviour (\ref{d1}) and the dynamics is described again by the action (\ref{W2}). 
%Therefore the theory is (almost) conformal with very weakly coupled spin-two fields $w_{\mu\nu}$.

%
%
%{\bf Question} How does the corresponding action look like? 
%It cannot be Weyl$^2$, since this also contains the massless graviton $w_{\mu\nu}$. Therefore most likely it is just Yang-Mills plus a free spin-two field $w_{\mu\nu}$.

\subsection{${\cal N}=4$ Super-Yang-Mills plus Super-Weyl theory}

\subsubsection{Massive theory}

%{\bf Include also ${\cal N}=4$, $U(N)$ super Yang-Mills part}

Now let us come to the ${\cal N}=4$ supersymmetric version of the Einstein, Yang-Mills plus Weyl$^2$ theory.
The spectrum of the ${\cal N}=4$ Super-Yang-Mills plus massive ${\cal N}=4$ Super-Weyl theory has the following form        \cite{Bergshoeff:1980is,Ferrara:2018wqd}:
    
\vskip0.5cm
\noindent
(i) A  standard massless spin-two super graviton multiplet $g_{{\cal N}=4}$ with $n_B+n_F=32$ degrees of freedom
and with the following helicities  and $SU(4)$ representations:
\begin{eqnarray}\label{gn4a}
   (+2,\underline1) + ( +{3\over 2} ,\underline 4) +(1,\underline 6) + ( +{1\over2},\underline{\overline4} )   + ( 0,\underline1)     \, ,
\end{eqnarray}
together with its CPT conjugate 
\begin{eqnarray} \label{gn4b}
  ( 0,\underline1)   + ( -{1\over2},\underline4 )  +(-1,\underline 6) 
   + ( -{3\over 2} ,\underline {\overline 4}) +(-2,\underline1) .
\end{eqnarray}
The complex scalar corresponds to the complex complex constant $\tau$ of the ${\cal N}=4$ field theory, i.e. to the massless marginal operator in the superconformal field theory.

\vskip0.5cm
\noindent
(ii) A  standard massless spin-one,  ${\cal N}=4$ super Yang-Mills  multiplet $W^a$ ($a=1,\dots ,N^2$) of the $U(N)$ gauge group with each $n_B+n_F=16$ degrees of freedom
and with the following helicities  and $SU(4)$ representations:
\begin{eqnarray}\label{ym4}
   (+1,\underline1) + ( +{1\over 2} ,\underline 4) +(0,\underline 6) + ( -{1\over2},\underline{\overline4} )   + ( -1,\underline1)     \, .
\end{eqnarray}
Here the $6\times N$ scalars from the Cartan subalgebra superfields are additional marginal operators, which parametrize the Coulomb branch of the ${\cal N}=4$
super Yang-Mills gauge theory. Giving them generic vev's breaks the $U(N)$ gauge symmetry to its maximal Abelian subgroup $U(1)^N$.
Together with the axion-dilaton field $\tau$ of the supergravity multiplet which couples to the quadratic YM action, these massless scalars parametrize the moduli space ${\cal M}$ of the theory which is given by the following coset space:
\begin{equation}
{\cal M}={SU(1,1)\over U(1)}\otimes R^{6N}. \label{r6n}
%{SO(6,N)\over SO(6)\times SO(N)}\, . 
\end{equation}
 Note  that the $6N$ scalars $\Phi_{ij}=-\Phi^{ji}, ~(i,j=1,\cdots,4),
 % ~(I=1,\cdots,N)
 $  of the $N$  vector multiplets are coupled to the curvature scalar in confrormal supergravity as 
\begin{eqnarray}
 {\cal L}=\cdots- \frac{1}{12}{\rm Tr}\Big(\Phi_{ij}\Phi^{ij}\Big)\Big(R+\cdots\Big),
 \end{eqnarray}
% where $\eta_{IJ}$ is a diagonal matrix with $6$ enties $-1$ and $N\!-\!6$ enties $+1$. 
 %={\rm diag}(-1,\cdots,-1,1,\cdots,1)$ has $
Therefore, the conditions
\begin{eqnarray}
{\rm Tr}\Big(\Phi_{ij}\Phi^{ij}\Big)=-6,~~~ {\rm Tr}\Big(\Phi_{ij}\psi^{j}\Big)=0, \label{ff3}
\end{eqnarray}
where $\psi^{j}$ ate the gauginos, break superconformal dilatations and S-supersymmetry,  
lead to Poincar\'e supergravity and in this case the scalars 
%of the vector multiplets 
parametrize  the coset  \cite{DFer,deroo,deroo1,Butter}
\begin{eqnarray}
{SU(1,1)\over U(1)}\otimes {SO(6,N)\over SO(6)\times SO(N)}. \label{ff4}
\end{eqnarray}
In fact, the conditions (\ref{ff3}) are weaker than the constraints
\begin{eqnarray}
{\rm Tr}\Big(\Phi_{ij}\Phi^{kl}\Big)=-\frac{1}{2}\delta^k_{[i}\delta^l_{j]},~~~ {\rm Tr}\Big(\Phi_{ij}\psi^{k}\Big)=0, \label{ff33}
\end{eqnarray}
 imposed by the equations of motion of the scalars ${D^{ij}}_{kl}$ and the fermion ${\chi^{ij}}_k$, which we describe in section 2.4.3. These constraints allow to remove six vector multiplets in massless Einstein supergravity. 
Notice that in rigid supersymmetry, the Yang-Mills scalar manifold is flat $R^{6N}$ whereas in Poincar\'e supergravity  the coset is $SO(6,N)/SO(6)\times SO(N)$. It looks that in massive Weyl supergravity the scalar manifold is  $SO(6,N)/SO(N)$ because 15 scalars have not been Higgsed.  In other words the constraints (\ref{ff3}) and (\ref{ff33}) remove the $1$ and $20$ from $6\times6=1+20+15$ but do not remove the $15$.  
The  first constraint in (\ref{ff33}) coming   from the  $D$ scalars which appear linearly in Einstein supergravity, is just a contribution to the scalar potential in massive Weyl supergravity because  the $D$ scalars appear now quadratically in the Lagrangian.
Hence, the deformation of (\ref{r6n})  to (\ref{ff4}) is only true if the Weyl term is absent so that the 15 gauge fields of the superconformal multiplet are auxiliary and their equations
of motion produce the deformation from $R^{6N}$ to $SO(6,N)/SO(6)\times SO(N)$.
However if the Weyl action term is added, the 15 vectors are massive and propagating and the above coset is not reproduced.
Poincar\'e supergravity is the limit $M_p\to \infty$  while Weyl supergravity is the limit $M_p=0$.
What happen in between is a new theory we are describing.
The potential of this new theory is different from Poincar\'e supergravity
 %because the D and E scalars contribute to the scalar potential
and is strictly quartic in all scalar fields before imposing the constraints as we will see below.

\vskip0.2cm
\noindent
(iii)  In the non-standard sector we have 
the spin-two massive Weyl multiplet of ${\cal N}=4$, which is irreducible with $n_B+n_F= 2^8=256$
states in $USp(8)$ representations       \cite{Bergshoeff:1980is}:
 \begin{eqnarray}\label{wN=4}
w_{{\cal N}=4}:~&   
{\rm Spin}(2) + {\underline{8}}\times {\rm Spin} ( 3/2 )  + {\underline{27}}
\times {\rm Spin}(1) + {\underline{48}}\times {\rm Spin} (1/2) + {\underline{42}}\times {\rm Spin} (0)\, .
    \end{eqnarray}  
\vskip0.2cm
\noindent
Hence in summary, the ${\cal N}=4$ massive super-(Weyl)$^2$ gravity theory contains $n_B+n_F=288+16N$ degrees of freedom, where $N$ is the number of  physical vector multiplets. General massive multiplets in extended supersymmetry were discussed in \cite{Ferrara:1980ra}

%Note that the ${\cal N}=4$ Weyl multiplet $w_{{\cal N}=4}$
%has the same structure as  the  ${\cal N}=4$ energy momentum tensor $T$ (i.e. supercurrent), which is given as:
%\begin{equation}
%T_{{\cal N}=4}\simeq{\rm Tr}(W^a\bar W^b )\, .
%\end{equation}

Also note that in Einstein supergravity  constraints (\ref{ff3}) and (\ref{ff33}) are field constraints while in massive Weyl supergravity they are
VEV constraints (Higgs phase) since the six vector multiplets, which appear in the massless limit (see next section)
are in this case physical degrees of freedom. As we will now see, 
in massless Weyl supergravity these multiplets become unphysical gauge degrees of freedom since the 
massless Weyl action does not depend on compensators being superconformal invariant. So in massless Weyl supergravity coupled to Yang-Mills the moduli space is  that in eq.(\ref{r6n}).
The massive phase is obtained when six extra singlet compensating vector multiplets are introduced.

%
%Let us also note that in Einstein supergravity the  constraints
% (\ref{ff3}) and (\ref{ff33})  are field constraints while in massive Weyl supergravity they are
%VEV constraints(Higgs phase) since the six vector multiplets are in this case physical degrees of freedom.
%In massless Weyl supergravity these multiplets become unphysical gauge degrees of freedom since the massless Weyl action does not depend on compensators being superconformal invariant. So in massless Weyl Sugra coupled to Yang-Mills, the moduli space is  that in eq.(\ref{r6n}).
%The massive phase is obtained when six extra singlet compensating vector multiplets are introduced.

\subsubsection{Massless theory}

Now we can consider the ${\cal N}=4$ supersymmetric version of the Higgs effect for the spin-two Weyl superfield $w_{{\cal N}=4}$.
In the limit $M_P\rightarrow 0$ the bosonic and fermionic degrees of freedom of $w_{{\cal N}=4}$ will arrange themselves into proper massless supermultiplets, when taking into
account the additional local superconformal and gauge symmetries, which arise in the massless limit.
In order to perform the massless limit we
 need  the branching rules of the massive $USp(8)$ R-symmetry group
 into the R-symmetry group $SU(4)$ of the massless states. The specific decomposition of $USp(8)\rightarrow SU(4)$ for the relevant representations is as follows:
\begin{eqnarray}
 {\underline{8}}&=&      { {\underline{{ 4}}}}\oplus{\overline {\underline{{ 4}}}}    \, ,  \nonumber\\
  {\underline{27}}&=&      { {\underline{{ 6}}}}\oplus{\overline {\underline{{ 6}}}}  \oplus{ {\underline{{ 15}}}}  \, ,  \nonumber\\
{\underline{42}}&=& { {\underline{{ 1}}}}\oplus{\overline {\underline{{ 1}}}}\oplus { {\underline{{ 10}}}}+{{\underline{{ {\overline {10}}}}}}\oplus{ {\underline{{ 20'}}}}\, , \nonumber \\
{\underline{48}}&=&   { {\underline{{ 20}}}}\oplus{\overline {\underline{{ 20}}}}  \oplus    { {\underline{{ 4}}}}\oplus{\overline {\underline{{ 4}}}}  
\end{eqnarray}

\vskip0.4cm\noindent
Then for $M_P=0$, the spectrum of the massless ${\cal N}=4$ Super-Weyl theory has the following form \cite{Ferrara:2018wqd}:

\vskip0.5cm
\noindent
(i) A  standard massless spin-two supergravity multiplet with $n_B+n_F=32$ degrees of freedom as given in eqs.(\ref{gn4a}) and (\ref{gn4b}).

\vskip0.2cm
\noindent
(ii)  In the non-standard sector,   we get first  from the massive Weyl multiplet  $w_{{\cal N}=4}$ a massless ghost-like spin-two supermultiplet  with $n_B+n_F=32$ and with the  helicites
and $SU(4)$ quantum numbers, again as given eqs.(\ref{gn4a}) and (\ref{gn4b}).

Second we get from $w_{{\cal N}=4}$ four massless spin-3/2 supermultiplets (in total $n_B+n_F=128$) with the following helicities  and $SU(4)$ representations, namely
\begin{eqnarray}\label{physa}
{{\underline{\bar{ 4}}}}\times\lbrack   ({3\over 2},\underline 1) +  ( 1,\underline 4 )+({1\over 2},\underline 6)+
( 0 ,\bar{\underline 4})+(-{1\over 2},\underline 1)
\rbrack  \, ,
\end{eqnarray}
together with the CPT conjugate states
\begin{eqnarray}\label{physb}
{{\underline{{ 4}}}}\times\lbrack   ({1\over 2},\underline 1) +  ( 0,\underline 4 )+(-{1\over 2},\underline 6)+
( -1 ,\bar{\underline 4})+(-{3\over 2},\underline 1)
\rbrack 
  \, .
\end{eqnarray}
They contain the 15 gauge bosons of the local $SU(4)_R$ gauge symmetry.

 In addition, the massive Weyl multiplet  $w_{{\cal N}=4}$ contains six ${\cal N}=4$ vector multiplets of the form:
  \begin{equation}\label{unphys}
 6~ ({\rm spin-one}):~  {{\underline{{  6}}}}\times\lbrack  
  (+1,\underline 1) +   (+{1\over2},\underline4 )
+( 0,\underline 6 )+(-{1\over2},\underline{\overline 4} )+(-1,\underline1).
\rbrack  
   \end{equation}
However these multiplets are unphysical since they can be gauged away by the superconformal transformations together with the local $SU(4)_R$ transformations.
Specifically, one of the 36 scalars in these vector multiplets is a Weyl mode.
Other  15 scalars are the helicity 
zero component of the massive vectors inside $w_{{\cal N}=4}$, which are gauged away by the local $SU(4)_R$ transformations.
%Moreover the six vectors correspond to the six central charges of the ${\cal N}=4$ algebra. However in the superconformal theory, these central charges are again not allowed. 
Hence all
six vector-multiplets are unphysical, do not propagate and get removed from the spectrum.

We should note that the dipole ghost graviton and the tripole ghost 
spin-$3/2$  sector are accompanied by a dipole ghost complex scalar since the action is a higher-derivative action. Indeed,  the equations of motion  are fourth-order for the spin-2 and third order for the spin-$3/2$ states. 
This fact is also discussed in \cite{Johansson:2018ues} at the Lagrangian level. 
This is not the case for the $SU(4)$ gauge bosons which have standard Yang Mills
action. 
The sugra higher derivative action also contains a singlet vector mode
which, together with the gauge bosons, is part of the higher derivative
gravitino action (which as pointed out above obeys third order equations of motion). In other words,  the cubic gravitino action simultaneously describes
the gravitino, the partner of the graviton, as well as the gravitini of the gravitino multiplet. 

\vskip0.2cm
\noindent
Hence, the massless ${\cal N}=4$ super-(Weyl)$^2$ gravity theory contains $n_B+n_F=192$ physical, propagating degrees of freedom.
The same spectrum  was also obtained in  \cite{Berkovits:2004jj} using the string twistor formalism for the construction of ${\cal N}=4$ super-(Weyl)$^2$ gravity. The spin $1/2$ have three sources, from 
the spin $3/2$ cubic gravitino kinetic term, the spin $1/2$ cubic kinetic term and the spin $1/2$ standard Majorana kinetic term.

At the end of this section, we can summarize the spectrum  of Weyl supergravity in the following way.
In pure Weyl supergravity without any
additional massless Yang-Mills multiplets, the six vector multiplets with 36=1+15+20 helicity zero  components
play the role of super-goldstone bosons.
%respectively the longitudinal components of the spin-two massive graviton (1), of the 15 spin-one
 %gauge bosons (Higgs phase) and the 20 scalars
 %which correspond to release the D20=0 constraint (of Poincare Sugra) 
 %and the 10+10*which are massive and propagating(from the graviton multiplets).
 In the massless conformal Weyl phase ($M_P\rightarrow 0$) the six compensators are not there and the spectrum goes from 256 massive + 32 massless states
 to 160+32=192 massless states. The 160=32+128 massless states correspond to the second graviton multiplet plus four gravitini multiplets.
On the contrary if we delete the Weyl square part and we keep the six compensator vector multiplets we have the constraints
(2.41) and (2.43),
and we get back massless spin-two Einstein supergravity.

%\newpage

\subsubsection{Scalar potential}

In this section we will consider some couplings between the Yang-Mills sector and the Weyl sector of the theory. In particular we will discuss the potential of the scalar fields that appear
in the ${\cal N}=4$ Yang-Mill and Weyl supermultiplets.
The scalar fields of the Weyl and the Yang-Mills multiplet of the
 ${\cal N}=4$ conformal supergravity\footnote{We use freely the terms Weyl and conformal supergravity in an interchangable way, and similalry for the terms Einstein and Poincar\'e supergravity.} coupled to super Yang-Mills transform under specific representations of  $SU(4)$ which are tabulated in table 1, where also their Weyl weights and chiral $U(1)$ weights $w$ and $c$, respectively are given \cite{Bergshoeff:1980is,deroo,deroo1,Butter}. The indices $i,j,\ldots$ and $a,b,\ldots$ are $SU(4)$ and $SU(1,1)$  indices, respectively. 
\begin{table}[h!]
\centering
 \begin{tabular}{c c c c} 
 \hline
 Scalars & SU(4) rep. & w & c \\ [0.5ex] 
 \hline\hline
 $\phi^\alpha$&{\underline{1}}&0&1\\
 $E_{ij}$ & {\underline{10}} & 1 & -1 \\ 
 %${T_{ab}}^{ij}$ & {\underline{6}} & 1 & -1 \\
 ${D^{ij}}_{kl}$ & {\underline{20}} & 2 & 0 \\
 $\Phi_{ij}$ & {\underline{6}} & 1 & 0 \\ [1ex] 
 \hline
 \end{tabular}
 \caption{Scalars of the Weyl multiplet ($\phi,E,D)$ and the Yang-Mills multiplet ($\Phi)$, together with their $SU(4)$ assignments, Weyl ($w$) and chiral ($c$) weights. }
\end{table}
In particular, $\phi^\alpha$ represent two-degrees of freedom associated to the $SU(1,1)/U(1)$ coset of the spin-two dipole ghost multiplet,
$E_{ij}$ is symmetric, ${D^{ij}}_{kl}$ is pseudoreal and $\Phi_{ij}$ is antisymmetric, and  in the adjoint representation of the gauge group $G$. They satisfy the relations
\begin{eqnarray}
&&\phi^\alpha\phi_\alpha=1, ~~~ E_{ij}=E_{ji}, ~~~
%{T_{ab}}^{ij}=-\frac{1}{2}{\epsilon_{ab}}^{cd}
%{T_{cd}}^{ij}, ~~~{T_{ab}}^{ij}=-{T_{ba}}^{ij}=-{T_{ab}}^{ji},
%\nonumber \\
%&&
~~~{D^{ij}}_{kl}=\frac{1}{4}{\epsilon^{ij}}_{mn}{\epsilon_{kl}}^{pq}
{D^{mn}}_{pq}, ~~~{D^{ij}}_{kj}=0,~~~ \Phi_{ij}=-\Phi_{ji},
\end{eqnarray}
whereas their complex conjugate fields are 
\begin{eqnarray}
&&\phi_1=(\phi^1)^*, ~~~\phi_2=-(\phi^2)^*,~~~E^{ij}=(E_{ij})^*, \nonumber \\
&&{D_{ij}}^{kl}=
({D^{ij}}_{kl})^*={D_{ij}}^{kl}, ~~~\Phi^{ij}=(\Phi_{ij})^*=
-\frac{1}{2}\epsilon^{ijkl}\Phi_{kl}. 
%~~~T_{cdij}=({T_{cd}}^{ij})^*
\end{eqnarray}
Notice that in Eq.(\ref{wN=4}) we have seen that the spin-two massive Weyl multiplet of ${\cal N}=4$ in the non-standard sector has  $n_B+n_F= 2^8=256$ states  which are arranged in $USp(8)$ representations as follows 
\begin{eqnarray}\label{wN=44}
%w_{{\cal N}=4}:~&   
{\rm Spin}(2) + {\underline{8}}\times {\rm Spin} ( 3/2 )  + {\underline{27}}
\times {\rm Spin}(1) + {\underline{48}}\times {\rm Spin} (1/2) + {\underline{42}}\times {\rm Spin} (0)\, .
    \end{eqnarray}  
Therefore the scalars in the massive multiplet are in the $ {\underline{42}}$ representation of $USp(8)$. The latter is decomposed under $SU(4)\subset USp(8)$ as 
\begin{eqnarray}
{\underline{42}}={\underline{20}}+{\underline{10}}+{\underline{\overline{10}}}+{\underline{1}}+{\underline{\overline 1}},
\end{eqnarray}
and it is associated to the pseudoreal ${D_{ij}}^{kl}~
 ({\underline{20}})$, the complex $E_{ij}~({\underline{10}}+
 {\underline{\overline{10}}})$ and the complex $\phi^\alpha~({\underline{1}}+
 {\underline{\overline{1}}})$ of table 1. The six scalars 
 $\Phi_{ij}(=-\Phi_{ji})~\mbox{in the}~{\underline{6}}~\mbox{of}~SU(4)$ and in the adjoint of the gauge group  are just the scalars of the Yang-Mills multiplet.
 Note that the fields ${D_{ij}}^{kl}~
 ({\underline{20}})$, which appear in the unphysical vector multiplets in eq.(\ref{unphys}),
 are unphysical in the massless limit. Moreover the scalars in the ${\underline{6}}+{\underline{\overline{6}}}$
 representations of the spin-3/2 multiplets (see eqs.(\ref{physa}) and (\ref{physb})) are not part of the scalar potential, because they originate from the graviphoton fields.

The most general Lagrangian for the ${\cal N}=4$ conformal supergravity has been constructed in \cite{Butter}. It turns out that it is completely specified by a single holomorphic and homogeneous of zeroth degree function ${\cal H}(\phi^\alpha)$ of the coset variables $\phi^\alpha$. 
The structure of the scalar potential for  ${\cal N}=4$ super Yang-Mills is coupled to ${\cal N}=4$ conformal supergravity can be read off from refs\cite{deroo,deroo1,Butter} and it turns out to be (in the notation of 
\cite{Butter})
\begin{eqnarray}
V&=& {\cal H}\left(\frac{1}{8}{D^{ij}}_{kl}{D^{kl}}_{ij}-\frac{1}{16}
E_{ij}E^{jk}E_{kl}E^{li}+\frac{1}{48}\Big(E_{ij}E^{ij}\Big)^2\right)\nonumber \\
&&+
\frac{1}{16}{\cal DH}{D^{ij}}_{kl}E_{im}E_{jn}\epsilon^{klmn}+
\frac{1}{384}{\cal D}^2{\cal  H}E_{ij}E_{kl}E_{mn}E_{pq}\epsilon^{ikmp}\epsilon^{jlnq}
-\frac{1}{48}E_{ij}E^{ij}{\rm Tr}\Big(\Phi_{kl}\Phi^{kl}\Big)
\nonumber \\
&&
+
\frac{1}{8}{D^{ij}}_{kl}{\rm Tr}\Big(\Phi_{ij}\Phi^{kl}\Big)
+\frac{1}{3}\overline{f}(\phi)
 E^{ij}{\rm Tr}\Big(\Phi^{kl}[\Phi_{ik},\Phi_{jl}]\Big)\nonumber \\
&&
+\frac{1}{4}|f(\phi)|^2{\rm Tr}\Big([\Phi_{ik},\Phi^{kj}][\Phi_{jl},\Phi^{li}]\Big)+h.c.,  \label{pp}
\end{eqnarray}
where  ${\cal D}$ is the operator 
\begin{eqnarray}
{\cal D}=-\phi^\alpha\epsilon_{\alpha\beta}\frac{\partial }{\partial\phi_\beta}, ~~~\mbox{and}~~~f(\phi)=\phi^1+\phi^2.
\end{eqnarray}
In  rigid supersymmetry, only the last term of  the potential (\ref{pp})
exists. All the other terms arise from the Weyl multiplet (terms proportional to ${\cal H}$ and its derivatives) and the gauge-matter coupling.  
Note  also that with the $U(1)$ charge $c$ assignment  $c({\cal H})=0$, $c({\cal DH})=2$ and $c({\cal D}{\cal H})=4$, the potential (\ref{pp}) is $U(1)$ invariant ($c(V)=0$)  since 
$c(E)=-1$, $c(D)=c(\Phi)=0$ and $c(\phi^\alpha)=1$.
Therefore the potential in eq.(\ref{pp}) is what we would call
``massless Weyl supergravity coupled to matter"
whose massive Poincar\'e  supergravity deformation is obtained by adding six compensator vector multiplets with constraints given as in eq.(\ref{ff3}). 

%
% "massive Weyl supergravit coupled to matter" and is very different from Poincare supergravity because of the first five terms and E not being auxiliary anymore.
%Before eq.2.56 we should have “massless Weyl supergravity coupled to matter” 
%whose massive Poincare sugradeformation is obtained by adding six compensator vector multiplets with constraints given as in 2.41 and 2.43

 The scalars ${D^{ij}}_{kl}$ are auxiliaries and can be integrated out leading to 
 \begin{eqnarray}
 V&=&{\cal H}\left(-\frac{1}{16}
E_{ij}E^{jk}E_{kl}E^{li}+\frac{1}{48}\Big(E_{ij}E^{ij}\Big)^2\right)-\frac{1}{128{\cal H}}\Big( {\cal D H} E_{im}E_{jn}\epsilon^{klmn}+2 {\rm Tr}\big(\Phi_{ij}^{kl}\big)
 \Big)^2\nonumber \\
&&+
\frac{1}{384}{\cal D}^2{\cal  H}E_{ij}E_{kl}E_{mn}E_{pq}\epsilon^{ikmp}\epsilon^{jlnq}-\frac{1}{48}E_{ij}E^{ij}{\rm Tr}\Big(\Phi_{kl}\Phi^{kl}\Big)
\nonumber \\
&&+\frac{1}{3}\overline{f}(\phi)
 E^{ij}{\rm Tr}\Big(\Phi^{kl}[\Phi_{ik},\Phi_{jl}]\Big)+\frac{1}{4}|f(\phi)|^2{\rm Tr}\Big([\Phi_{ik},\Phi^{kj}][\Phi_{jl},\Phi^{li}]\Big)+h.c.\, , \label{intD}
 \end{eqnarray}
 where 
 \begin{eqnarray}
 \Phi^{ij}_{kl}=\Phi^{ij}\Phi_{kl}-2\delta^{[j}_{[l}\Phi^{i]m}\Phi_{k]m}+\frac{1}{3} \delta^i_{[k}\delta^j_{l]}\Phi^{pq}\Phi_{pq}.   
 \end{eqnarray}
 Note that a non-constant ${\cal H}$  function gives extra terms to the scalar potential (\ref{intD}).
This will be the case in twistor string theory  where ${\cal H}$ is an exponential in the holomorphic variable \cite{Berkovits:2004jj}.
For constant ${\cal H}$, the terms proportional to ${\cal DH}$ and ${\cal D}^2{\cal H}$ in the potential drop and it is easy to see that $E=0$ and $\Phi$ in the Cartan  subalgebra
of the gauge group is an extremum of the potential. This is the breaking of superconformal to Poincare supergravity if 6 auxiliary vector multiplets are added
with wrong sign so that a correct Einstein term and the solution $D=0$ is possible. Indeed, let us recall  that the fermions of the  theory are the gravitini $\psi_\mu^i$ (in the $4$ of $SU(4)$) associated with Q-supersymmetry, the composite $\phi_{\mu i}$ (in the $\overline{4}$) associated with S-supesymmetry and the two spinor fields  
$\Lambda_{i}$ and ${\chi^{ij}}_k$ in the $\overline{4}$ and $20$ of $SU(4)$, respectively. The fermionic shifts of the spinors fields contain  among others, the terms \cite{Bergshoeff:1980is}
\begin{eqnarray}
\delta \Lambda_i&=&\cdots+E_{ij}\epsilon^j,\nonumber \\
\delta {\chi^{ij}}_k&=&\cdots+{D^{ij}}_{kl}\epsilon^l-\frac{1}{2}
\epsilon^{ijlm}E_{kl}\eta_m ,\label{shift}
\end{eqnarray}
where $\epsilon^i$ and $\eta^i$ are the Q- and S-supersymmetry parameters. Therefore,  $E=0$ and $D=0$ are the necessary conditions for unbroken supersymmetry. In addition, for Poincar\'e supersymmetry, breaking of Weyl symmetry is required. This is achieved by imposing the condition (\ref{ff3}) while still $E=D=0$. 
If there are non-trivial extrema of the scalar potential beyond the supersymmetric Poincar\'e one is an interesting open problem. Such vacua will further break Poincar\'e supersymmetry, which will happen if the  $E$ and $D$ scalars have  non-vanishing vev.
 
We note that  pure massive Weyl supergravity is obtained by adding to the Weyl multiplet  6 vector multiplets of wrong sign.  
In this case the spectrum is the standard massless ${\cal N}=4$ Poincar\'e supergravity coupled to a massive ${\cal N}=4$  spin-2 ghost  multiplet.
The massive scalars are then  20 from the six compensatos, the $10+\overline{10}$ $E$ scalars and $1+1$ from the dilation dipole massive ghost. All together they make the $42$ (of $USp(8)$) as it should. Indeed, the constraint (\ref{ff33}) is needed in Poincare supergravity because the $D$ scalars appear linearly in the action
\cite{deroo,deroo1}. However, this is not true  in Weyl massive supergravity  where they appear quadratically \cite{Butter} so that they lead to a new    potential term  after integrate them out rather than to a  constraint and the $20$ D scalars becomes dynamical. In the higgsed phase,  the $15$ scalars go away
and this explains $42=1+1+20+10+\overline{10}$.

\section{String realization}

Now we want to discuss how to obtain Weyl$^2$ gravity plus Yang-Mills gauge theory from IIB superstring theory.
As already mentioned, in string theory the graviton $g_{\mu\nu}$ originates from the closed string sector and lives in the bulk space, whereas the 
 spin-two field $w_{\mu\nu}$ as well as the Yang-Mills gauge bosons $A_{\mu}^a$ come from the open string sector and will be localized on the world-volume of a stack
 of D3-branes. In the following we will first discuss the closed and open string spectrum and then, how the various limits can be realized in string theory.

 %\subsection{Spectrum}
 
 Here we will discuss the case of maximal supersymmetry. This means that in four-dimensional the closed string bulk theory possesses ${\cal N}=8$ supersymmetry (i.e. 32 supercharges),
 whereas the open string sector localized on the D-brane worldvolume will preserve ${\cal N}=4$ supersymmetry (i.e. 16 supercharges).
 Specifically, we will consider the type IIB superstring on $R^{1,3}\times T^6$, with an additional stack of N D3-branes with world-volumes on $R^{1,3}$. Possible other D-branes and/or orientifold
 planes do not play an important role for the discussion, and we also do not address the question of tadpole cancellation. In fact, when taking the decoupling limit of infinite $T^6$ volume
 later on, i.e. considering a
 non-compact six-dimensional extra space, 
we just  deal with N D3-branes in flat ten-dimensional space-time.

The spectrum of this string theory is now as follows:

\subsection{Open string sector}

\subsubsection{Massless open string Yang-Mills sector}

Now we come to the massless open string spectrum of the D3-branes on the background $R^{1,3}\times T^6$. 
For maximally supersymmetric, toroidal compactifications of $D=10$ superstring,
its excitations form supermultiplets of ${\cal N}=4$ supersymmetry. Before
discussing the first excited level, we recall
the vertices of massless particles, which arise from the zero modes and
include, in
the NS sector, the gauge bosons $A^a$ and six real scalars $\phi^I, ~I=1,\dots,
6.$ In the R sector, we have four gauginos $\lambda^A, ~I=A,\dots, 4$. All
in all, these zero mode form one ${\cal N}=4$
gauge supermultiplet.
The NS sector vertices, in the $(-1)$-ghost picture, read:
\begin{eqnarray} \label{onepic}
V_{A^a}^{(-1)}(z,\epsilon,k) &=& g_A\ T^a\ e^{-\phi}\
\epsilon^{\mu}\, \psi_{\mu}\ e^{ikX}\ ,\cr
V_{\phi^{a,I}}^{(-1)}(z,k) &=& g_A\ T^a\ e^{-\phi}\ \Psi^{I}\
e^{ikX}.\end{eqnarray}
Here, $X,\psi,Z,\Psi$ are the fields of ${\cal N}=1$ worldsheet SCFT, with the Greek
indices associated to $D=4$ spacetime fields $X^\mu, \psi^\nu$ and the Latin
upper case labeling internal $D=6$ (e.g. $Z^I, \Psi^I$). $\phi$ is the scalar
bosonizing the superghost
system.

The R sector vertices, in the $(-1/2)$-ghost picture, read:
\begin{eqnarray}\label{vertlam}
V_{\lambda^{a,A}}^{(-1/2)}(z,u,k)
&=&g_\lambda\ T^a\ e^{-\phi/2}\ u^{\sigma} S_{\sigma}\ \Sigma^A\ e^{ik X}\ ,\cr
V_{\bar\lambda^{a,A}}^{(-1/2)}(z,\bar u,k)
&=&g_\lambda\ T^a\ e^{-\phi/2}\ \bar u_{\dot\sigma} \bar S^{\dot\sigma}\
\overline\Sigma^A\
e^{ikX}.\end{eqnarray}
Here, $S$ and $\bar S$ are the left and right-handed $SU(2)$ spin fields,
respectively, while $\Sigma^A$ and $\overline\Sigma^A$ are the internal Ramond spin fields.
%discussed in the following Subsection.
The couplings are
\begin{equation}
g_A=(2\alpha')^{1/2}\ g_{YM}\quad,\quad g_\lambda=(2{\alpha'})^{1/2}{\ap}^{1/4}\ g_{YM}\ ,
\end{equation}
where $g_{YM}$ is the gauge coupling.
In the above definitions, $T^a$ are the Chan-Paton factors accounting for the
gauge
degrees of freedom of the two open string ends, meaning that all these massless states are in the adjoint representation of the $U(N)$ gauge group.

We can also write these states in terms of the fermionic oscillators in the transversal  space-time directions, denoted by  $b_r^i$ ($i=1,2$), and the internal oscillators $b_r^I$ ($I=1,\dots, 6$).
Then the   eight bosons bosonic states in he adjoint representation look like
\begin{equation}
A_i^a\,\sim\, T^ab_{-1/2}^i|0\rangle\, ,\quad \Phi^{a,I}\,\sim\, T^ab_{-1/2}^I|0\rangle
\end{equation}
For the eight fermions in the adjoint representation one simply has
\begin{equation}
\lambda^{a,A}\,\sim\, T^a
|\dot \alpha,A\rangle\, ,
\end{equation}
where $|\dot \alpha,A\rangle$ is the Ramond ground state with four-dimensional spinor-helicity index $\dot\alpha=1,2$ and internal spinor index $A=1,\dots ,4$.
These states indeed built massless ${\cal N}=4$ vector multiplets in the adjoint representation of the gauge group $U(N)$, which are
displayed in eq.(\ref{ym4}).
They are  localized at the world-volume of the N D3-branes.

%See paper with Oliver

\subsubsection{Massive open string spin-two sector}

%Square of Yang-Mills

We will now determine the first excited, massive open string states, which are also localized at the world-volume of the N D3-branes. 
For maximally supersymmetric, toroidal compactifications of $D=10$ superstring,
NS and R sectors form one spin-two  massive supermultiplet of ${\cal N}=4$
supersymmetry. The bosons form one symmetric tensor field $B_{mn}$ and one
completely antisymmetric tensor field $E_{mnp}$. Here, the indices $(m,n,p)$
label $D=10$. All these particles are in the adjoint representation of the
gauge group.
The corresponding vertices, in the $(-1)$-ghost picture, read
\cite{Feng:2010yx}:
\begin{equation}\label{verteb}
V_{N\! S,a}^{(-1)}(z,k) = \frac{g_A}{\sqrt{2\alpha'}}\ T^a\ e^{-\phi}(\,
E_{mnp}\,\psi^m\psi^n\psi^p \,
+
\, B_{mn}\, i\partial X^m\psi^n \, + \, H_m\partial\psi^m \,
)\,e^{ikX}\ ,
\end{equation}
where $H_m$ is an auxiliary vector field. Note that again the open string gauge coupling 
$g_A=(2\alpha')^{1/2}\ g_{YM}$
appears in this vertex operator.
At this level, the on-shell
condition is $k^2=-\frac{1}{\ap}$. The constraints due to the requirement of BRS
invariance are:
\begin{eqnarray}\nonumber
~~~~k^mE_{mnp}&=&0\ ,\\
2\ap k^mB_{mn} + H_n &=&0\ ,\label{brstbos}\\
B^m_m+k^mH_m&=&0\ .\nonumber
\end{eqnarray}
In $D=10$ all 128 bosonic degrees of freedom can be accounted for by setting
$H=0$,
{\em i.e.} with a traceless, transverse $B$ and transverse $E$. 

Also for the fermions, we begin with the first massive level in $D=10$. In the R
sector, the fermion vertex operator [in its canonical $(-1/2)$-ghost picture] is
parametrized by two vectors, Majorana-Weyl spinors $v_m^A$ and $\bar
\rho^n_{\dot B}$ of opposite chirality \cite{Feng:2010yx}:
\begin{equation}
V_{R,a}^{(-1/2)}(z,v, \bar{\rho},k) \ \ = \ \ C_{\Lambda} \  T^a \  \bigl[
\,  v_m^{A} \  i\partial X^m \ + \ 2\alpha' \, \bar{\rho}^m_{\dot B} \ \psi_m\, \psi^n
\, \Gamma_n^{\dot B A} \, \bigr] \,  \Theta_{A}  \  e^{-\phi/2} \ e^{ik  X} \ .
\label{massR}
\end{equation}
Here, $A$ denotes a left-handed spinor index while $\dot B$ is its right handed
counterpart. $\Gamma_n$ are $16 \times 16$ Weyl blocks of the $D=10$ gamma matrices
and $\Theta_A$ are the conformal weight $h=\frac{5}{8}$ chiral spin fields.

{}Requiring BRST invariance imposes two on-shell constraints on $v_m^A$ and
$\bar \rho^m_{\dot B}$ which determine $\bar \rho$ in terms of $v$ and leave 144
independent components in the latter. Furthermore, a set of 16 spurious states
exists which allows to take $\bar \rho$ and $v$ as transverse and
$\Gamma$-traceless:
\begin{equation}
k^m \, v_m^A \eq v_m^A \, \Gamma^m_{ A \dot B} \eq k_m \, \bar \rho^m_{\dot B} \eq
\bar \rho^m_{\dot B} \, \Gamma_m^{\dot BA} \eq 0\ .
\end{equation}
These $128=144-16$ physical degrees of freedom match the counting for bosons.

As for the massless states, we can also write the massive states, that are created by these vertex operators, in terms of the bosonic and fermionic oscillators $\alpha_n$ and $b_r$.
Now we split the indices into uncompactified and internal indices.
Furthermore we will omit the gauge index, i.e. we drop the Chan-Paton factor $T^a$, which means that we consider the neutral, excited states of the Abelian $U(1)$ vector-multiplet.
This $U(1)$ gauge group is just the Abelian part of the full gauge group $U(N)=SU(N)\times U(1)$. Alternatively we could consider the case of a single D3-brane, i.e. $N=1$, where
the excited states are also neutral.
Then one  obtains at the first massive level the following massive open string states (see for example \cite{Blumenhagen:2013fgp}):
\begin{eqnarray}\label{bosons2}
 &~&b^i_{-1/2}b^j_{-1/2}b^I_{-1/2}|0\rangle\, ,\qquad b^i_{-1/2}b^I_{-1/2}b^J_{-1/2}|0\rangle\, ,\qquad b^I_{-1/2}b^J_{-1/2}b^K_{-1/2}|0\rangle\, ,\nonumber\\
&~&b^i_{-3/2}|0\rangle\, ,\qquad   b^I_{-3/2}|0\rangle\,   \nonumber\\
&~& \alpha_{-1}^ib^j_{-1/2}|0\rangle\, ,\quad \alpha_{-1}^ib^I_{-1/2}|0\rangle\, ,\quad\alpha_{-1}^Ib^i_{-1/2}|0\rangle\, ,\quad\alpha_{-1}^Ib^J_{-1/2}|0\rangle\, .
\end{eqnarray}
(Here the $b$'s and the $\alpha$'s are the oscillators of the world-sheet fermions and bosons.)
Collecting all states and putting them into proper massive representations of the four-dimensional little group $SO(3)$ as well as in 
proper representations of the ${\cal N}=4$   $SU(4)$ R-symmetry, one obtains the following massive spectrum:
\begin{equation}\label{N4bos}
{\underline 1}\times {\rm Spin}(2)+({\underline 6}+{\underline 6}+{\underline {15}})\times {\rm Spin}(1)+(2\times\underline1+{\underline {10}}+{\bar{\underline {10}}}+{\underline {20'}})\times {\rm Spin}(0)\, .
\end{equation}
For massive states in ${\cal N}=4$ supersymmetry the R-symmetry group is enhanced from $U(4)$ to $USp(8)\supset U(4)$ with the following branching rules:
\begin{eqnarray}
{\underline 8}&=&{\underline 4}+{\bar{\underline 4}}\, ,\nonumber\\
{\underline {27}}&=&{\underline 6}+{\underline {6}}+{\underline{15}}\, ,\nonumber\\
{\underline {36}}&=&{\underline 1}+{\underline {10}}+\bar{\underline {10}}+{\underline{15}}\, ,\nonumber\\
{\underline{42}}&=&2\times\underline1+{\underline {10}}+{\bar{\underline {10}}}+{\underline {20'}}\, ,\nonumber\\
{\underline{48}}&=&{\underline 4}+\bar{\underline 4}+{\underline {20}}+{\bar{\underline {20}}}
\end{eqnarray}
Then the massive bosons transform under $USp(8)$ as
\begin{equation}\label{N4bossp}
{\underline 1}\times {\rm Spin}(2)+({\underline {27}})\times {\rm Spin}(1)+({\underline{42}})\times {\rm Spin}(0)\, .
\end{equation}

In ten dimensions, the 128 massive fermions are given by the following string states:
\begin{eqnarray}\label{fermions1}
 (8)_c+(56)_c:\quad b^A_{-1}|a\rangle\, ,\qquad        (8)_s+(56)_s:\quad \alpha^A_{-1}|\dot a\rangle\,   \, .
\end{eqnarray}
In terms of four-dimensional massive spinors this leads to:
\begin{equation}
%\label{fermions2}
({\underline 4}+\bar{\underline 4})\times{\rm Spin}(3/2)+({\underline 4}+\bar{\underline 4}+{\underline {20}}+{\bar{\underline {20}}})\times {\rm Spin}(1/2)\, ,
\end{equation}
where in this decomposition each spin 3/2 Rarita Schwinger field in four dimensions contains 4 degrees of freedom and each  spin 1/2 Dirac fermion possess 2 degrees of freedom.
Under $USp(8)$ the massive fermions transform as
\begin{equation}\label{fermions2}
({\underline 8})\times{\rm Spin}(3/2)+({\underline {48}})\times {\rm Spin}(1/2)\, ,
\end{equation}

The bosons in eq.(\ref{N4bos}) together with the fermions in eq.(\ref{fermions2}) build one long, massive ${\cal N}=4$ spin 2 supermultiplet.
It precisely agrees with the super Weyl multiplet $w_{{\cal N}=4}$, which is displayed in eq.(\ref{wN=4}).
%\subsection{Field theory}

\subsection{Closed string  sector}

In the following we will also provide the closed string spectrum of the theory, both in the bulk and also on the stack of the D3-branes. The vertex operators are similar to one of the
open strings and obtained by the tensor product of left- and right-moving open string states at each mass level, taking into account the level matching constraint $h_L=h_R$.

\subsubsection{Massless gravity sector}

Let us us first recall the closed string type II B spectrum of the bulk theory on the background space $R^{1,3}\times T^6$. As it is well known, the massless closed string states originate
from the (NS,NS), (R,R), (R,NS) and (NS,R) sectors of the theory. Altogether they built the standard massless ${\cal N}=8$ supergravity multiplet with $n_B+n_F=256$ propagating massless
degrees of freedom. However on the world volume of the stack of N D3-branes supersymmetry is broken by half from ${\cal N}=8$ to ${\cal N}=4$, where 16 supersymmetries are linearly 
realized and the other half of 16 supersymmetries are non-linearly realized on the D3-branes. 
Therefore the massless closed string spectrum on the D3-branes is precisely the one of ${\cal N}=4$ supergravity.
The corresponding massless states precisely build the standard 
massless spin-two super graviton multiplet $g_{{\cal N}=4}$, which is displayed in eqs.(\ref{gn4a}) and (\ref{gn4b}).

\subsubsection{Massive closed string spin-four sector}

As discussed in \cite{Ferrara:2018iko}, the first excited closed string states are obtained by performing the tensor product of 
two super-Weyl supermultiplets. This leads to a massive supermultiplet with a highest spin-four tensor field  $\Phi^4$ in the closed string sector, whereas 
the massive spin-two sector, i.e. the massive Weyl supermultiplets, correspond to open string excitations.

  For the case under consideration with 
background space $R^{1,3}\times T^6$, the bulk spectrum is then given in terms of  massive spin-four ${\cal N}=8$ supermultiplet $ \Phi^4_{{\cal N}=8}$:
\begin{equation}
\Phi^4_{{\cal N}=8}=w_{{\cal N}=4} \otimes w_{{\cal N}=4} \, .
\end{equation} 
It contains $n_B+n_F=256 \times 256 =   10^{16}=65.536    $ degrees of freedom.
When restricting it to the world volume of the N D3-branes, it gets truncated and becomes massive spin-four ${\cal N}=4$ supermultiplet $ \Phi^4_{{\cal N}=4}$ with
$n_B+n_F= 1280$.
Its exact multiplet structure is as follows:
\begin{eqnarray}\label{spinfour}
&~&{\underline 1}\times {\rm Spin}(4)+
{\underline 8}\times{\rm Spin}(7/2)+
(\underline 1+{\underline {27}})\times {\rm Spin}(3)+
({\underline 8}+{\underline{48}})\times{\rm Spin}(5/2)
\nonumber\\
&~&
+(\underline1+{\underline {27}}
+
{\underline{42}})\times {\rm Spin}(2)+
({\underline 8}+{\underline{48}})\times{\rm Spin}(3/2)+
(\underline 1+{\underline {27}})\times {\rm Spin}(1)
\nonumber\\
&~&
+{\underline 8}\times{\rm Spin}(1/2)
+{\underline{1}}\times {\rm Spin}(0)\, .
\end{eqnarray}

%GIVE THE EXPLICIT MULTIPLET STRUCTURE.

 \subsection{Effective field theory and limits}
 
 Now we will discuss the four-dimensional  effective field theory on the stack N D3 branes. From the closed strings we will restrict ourselves to the massless
 gravitational sector, and  the closed string spin-four  in the bulk will be mentioned later in the next section on holography.
For the open strings, we will on the massless spin-one Yang-Mills sector as well as on the massive spin-two Weyl sector.
Since both types of fields belong to open string with ends lying on the D3-branes, the Yang-Mills field as well as the Weyl fields are confined to the world-volumes of the D3-branes.

 \subsubsection{Ten-dimensional picture, non-compact space}

Here we consider a stack of N D3-branes in a non-compact space $R^{1,9}$. 
The ten-dimensional action can be schematically written as
\begin{equation}
S=S_{\rm bulk}+S_{\rm brane}+S_{\rm int}\, ,
\end{equation}
where $S_{\rm bulk}$ is the effective action of the massless gravitons and their superpartners from the closed strings, 
$S_{\rm brane}$ is the four-dimensional effective action of the massless Yang-Mills fields and the massive spin-two field $w_{\mu\nu}$ on the D3-branes,
\begin{equation}
S_{\rm brane}=S_{YM}+S_W\, ,
\end{equation}
 and $S_{\rm int}$ describes the interactions between the open and closed string modes.

 \vskip0.5cm
 \noindent
 Let us now determine the effective couplings in terms of the basic string parameters, which are
 
 (i) $g_s=e^{\phi}$, the string coupling constant, which is determined by the vev of the dilaton and

(ii) $M_s=1/\sqrt {\alpha'}$, namely the string scale.

\vskip0.5cm
 \noindent
 In the string frame, the effective ten-dimensional Planck mass is given as
 \begin{equation}
\kappa^{(10)}=\Bigl(M_P^{(10)}\Bigr)^{-4}={1\over M_s^4}g_s\, .
\end{equation}
The masses $M_n$  of the string excitations in the string frame directly follow from the fundamental string tension and are given by $M_n^2=nM_s^2$. Namely in the string scale the mass $M_W$ of the
first open string excitations is simply given as
\begin{equation}
M_W=M_s\, .
\end{equation}  

In order to go to the Einstein frame, one has to perform a Weyl rescaling of the metric, which in D dimensions takes the form
\begin{eqnarray}
 &~&               g \rightarrow \exp(\phi/2) g\, ,\nonumber\\
 &~&\sqrt{|g|}_D \rightarrow \exp( D \phi/4 ) \sqrt{|g|}_D\, ,\nonumber\\
      &~&          R \rightarrow \exp(-\phi/2) R  \, .
                   \end{eqnarray} 
                (Hence for $D=4$  the Weyl action $W^2 \sqrt{|g|}$ is indeed invariant under this rescaling.)

Therefore the  ten-dimensional  Einstein-Hilbert term transforms from the string frame into the Einstein frame as 
\begin{eqnarray}
\sqrt{|g|}_{10} e^{-2 \phi} R   \rightarrow \sqrt{|g|}_{10}  R 
 \end{eqnarray} 
 and in the Einstein frame the Planck mass is therefore independent of $g_s$:
 \begin{equation}
\kappa^{(10)}=\Bigl(M_P^{(10)}\Bigr)^{-4}={1\over M_s^4}\, .
\end{equation}

Second, the gauge kinetic term of a 
 Dp-brane transforms from the string frame into the Einstein frame as
\begin{eqnarray}
\sqrt{|g|}_{p+1} e^{-\phi}   F^{\mu \nu}   F_{\mu \nu}  \rightarrow e^{( (p-7)/4 \phi)}  \sqrt{|g|}_{p+1}   F^{\mu \nu}   F_{\mu \nu}
\end{eqnarray}
Hence, for D3-branes ($p=3$) the effective gauge coupling in the Einstein frame is given as 
\begin{equation}
g_{YM}=\sqrt{g_s}\, .
\end{equation}

Finally for the  fundamental string tension one obtains that
\begin{eqnarray}
\sqrt{|g|}_{1+1}    \rightarrow  e^{\phi/2} \sqrt{|g|}_{1+1}   
\end{eqnarray}
Therefore the masses of the excited  strings in the Einstein frame scale as
 \begin{equation}
 M^2 _n\sim n\sqrt{g_s}M_s^2\, ,
 \end{equation}
  and hence the ratio  between these masses and the 
10d Planck scale remains invariant.
In D dimensions, a mass, when measured in the Einstein metric, is related to $g_s$ as
\begin{equation}\label{excited}
 M^2 _n\sim n{g_s}^{{4\over D-2}}M_s^2\, .
 \end{equation}

 In the limit $\alpha'=M_s^{-2}\rightarrow 0$, while keeping $g_s$, $N$ and all other physical 
 length scales, such as curvature scales fixed, all massive string excitations decouple 
  and the higher derivative interactions can be neglected. Furthermore, open and closed  string modes decouple and gravity becomes free, i.e. we arrive at a theory of free 
  gravitons and its supersymmetry partners.  This decoupling limit  is  also referred to the Maldacena limit: free type IIB supergravity in the bulk and four-dimensional SYM theory with 16 supercharges on the world-volume of the branes.
   % In addition the spin-two field $w_{\mu\nu}$ become infinitely heavy and also completely decouple.
  To  see the more precise form of the decoupling limit, which zooms into the near horizon region of the D3-brane SUGRA solution, we recall that it is defined as follows:
\begin{equation}
L\,M_s\,\rightarrow\, \infty\quad{\rm with}\quad L^4={g_sN\over M_s^4}\, .
\end{equation}
On the gauge theory side this limit corresponds to the limit of infinite 't Hooft coupling 
\begin{equation}
\lambda\,\rightarrow\, \infty\quad{\rm with}\quad \lambda=g_{YM}^2N\, .
\end{equation}
In this the near-horizon limit the type IIB background of the N D3 branes is given by the well-known $AdS_5\times S^5$ geometry.
Note that this limit can be obtained by sending $L$ to infinity while keeping $M_s$ fixed, which means that the near horizon limit can be obtained for finite masses of the 
string excitations.

%Therefore, in this limit also the massive spin-two field $w_{\mu\nu}$ becomes very weakly coupled.

\subsubsection{Four-dimensional picture, compact internal space}

Now we switch to four dimensions and consider the theory  compactified 
 on $R^{1,3}\times T^6$.  
 As we have discussed in section 3.1.2 the  massive open string excitations precisely agree with the ${\cal N}=4$ spin-two Weyl supermultiplet. 
 The question is now, which is the correct effective action for these massive states. Since these states appear at the first mass level,
 the corresponding effective action must contain four derivatives. Hence a priori, it could be either the $R^2$-action or the $W^2$-action.
 Since the $R^2$ propagates a scalar degree of freedom, whereas the  $W^2$-action propagates precisely the spin-two degrees of freedom of the Weyl-supermultiplet,
 we can safely conclude that the $W^2$-action is the correct effective action for the massive open string fields.
 Therefore the four-dimensional  string effective active action for closed string gravity plus open string Yang-Mills plus open string massive spin-two fields has the following form:
 \begin{eqnarray}
S_{\rm eff}= \int d^4 x\sqrt{-g}\Big{(}-{1\over 4 g_{YM}^2}  F_{\mu\nu}^aF^{a\,\mu\nu}+{1\over 2 g_W^2}  W_{\mu\nu\rho\sigma}W^{\mu\nu\rho\sigma}+M_P^2 R\Big{)}\, .
\end{eqnarray}
%On the string side, we have the  following four-dimensional  parameters resp. coupling constants:

%(i) $g_s=e^{\phi}$: the string coupling constant, which is determined by the vev of the dilaton.

%(ii) $M_s=1/\sqrt {\alpha'}$: the string scale, i.e. the mass of the string excitations.
In addition to the ten-dimensional string parameters $g_s$ and $M_s$ we now gain a third parameter, namely:

(iii) $R$: the radius of the internal space, i.e. the volume of the $T^6$ is given by $R^6$. In units of the string length $L_s$ the size of the internal scape is given by the dimensionless
parameter
$r=R/L_s=R M_s$.

The three string parameters $g_s$, $M_s$ and $r$ are identified with the three four-dimensional coupling constants of the effective theory in the following way:

(i) The four-dimensional gravitational closed string  coupling $M_P$ in the Einstein frame:

\begin{equation}
M_P=M_sr^3\, .
\end{equation}

(ii) The open string Yang-Mills coupling $g_{YM}$ for the gauge fields on the D3-branes:

\begin{equation}
g_{YM}=\sqrt{g_s}\, .
\end{equation}

(iii) The  bimetric Weyl coupling $g_{W}$:

The effective 4D coupling  $g_{W}$ can be determined by the requirement that the mass of the massive open string spin-two fields $w_{\mu\nu}$ 
 is given in the Einstein frame as (see eq.\ref{excited}))
\begin{equation}
M_W={{g_s}}\,M_s\, .
\end{equation}
%Note that the mass of the open string excitations on the D3-brane differs from the mass of the closed string excitation in eq.(\ref{excited}) by a factor of $g_s^{1/2}$, just as the open and closed
%string gauge couplings also differ by the same factor in the string coupling constant.
It then follows from eq.(\ref{mw}) that
\begin{equation}
g_{W}={g_s}/r^3\, .
\end{equation}
Observe that in the four-dimensional Einstein frame, the Weyl coupling $g_W$ is scaling with respect to $g_s$ as the gravitational  coupling, 
because it corresponds to a coupling between closed and open strings.
Moreover
is
proportional to the inverse of the internal volume.
%, whereas in the string frame $g_W$ is independent of $r$.
%{\bf This still needs some better understanding.}

Now we can  consider the following four decoupling limits in the four-dimensional effective string theory, which we already mentioned before in section two:

\vskip0.2cm
\noindent{\sl (A) Decoupling of gravity}
\vskip0.2cm

\noindent
The decoupling of the closed string modes namely the decoupling of standard gravity is achieved sending the Planck mass to infinity:
\begin{equation}
M_P \,\rightarrow\, \infty\, .
\end{equation}
In this limit  either the string scale $M_s$ is very large, i.e. $\alpha'\rightarrow 0$ with $r$ kept fixed.
 %which means that the size of the internal space is of the order of the string length: $R\simeq L_s$.
Alternatively one can keep $M_s$ finite, but sending $r\rightarrow\infty$, implying that $R>>L_s$ and the internal space becomes very large. Then the near horizon geometry close to the
N D3-branes becomes $AdS_5\times S^5$. In this sense the size $R$ of the internal space corresponds to the length parameter $L$ in the non-compact case.
Both, for finite $r$ and large $M_s$ and also for large $r$ and finite $M_s$ the massive spin-two open string fields $w_{\mu\nu}$ decouple, because these fields become either
very heavy ($M_s$ large) or their coupling constant $g_W$ becomes very small ($r$ large).

\vskip0.2cm
\noindent{\sl (B) Massless bigravity limit}
\vskip0.2cm

\noindent
Second, we consider  the massless limit, namely the limit of vanishing Planck mass: 
\begin{equation}
 M_P\rightarrow 0\, .
\end{equation}
It can be realized in string theory by sending the string scale $M_s$ to zero:  $M_s\rightarrow 0$ or equivalently $\alpha'\rightarrow\infty$.
In this limit, the open string spin-two fields become massless and the bimetric gravity theory becomes conformal.
However in string theory this is the tensionless limit, where an infinite tower of string states becomes massless in this limit.
Therefore the massless bimetric gravity theory only exists as an enormous truncation of higher spin theory with an infinite number of massless higher spin fields.
Alternatively, for fixed string scale $M_s$,  a vanishing Planck mass is obtained by sending $r\rightarrow0$. Here the size of the internal space becomes much
smaller than the string length. Furthermore $g_W$ becomes large and the open string spin-two fields $w_{mu\nu}$ become strongly coupled.
%Finally, $M_P$ can be made small for large string coupling, i.e. $g_s\rightarrow\infty$. This limit has the disadvantage that the string and also the open string spin-two fields $w_{\mu\nu}$ become strongly coupled.

%\vskip0.2cm
%\noindent{\sl (ii) Einstein plus Yang-Mills limit}
%\vskip0.2cm

%Now let us consider the case Here  MP fixed,    Ms finite  i.e. r large

%free spin-two decoupled

\vskip0.2cm
\noindent{\sl (C) Light spin-two plus massless Yang-Mills limit}
\vskip0.2cm

\noindent
Now let us consider the case where the string scale is very small compared to the Planck mass. This is the socalled  low string scale scenario, which implies large extra dimensions:
\begin{equation}
M_s << M_P\, .
\end{equation}
This limit can be achieved by sending $r\, \rightarrow\, \infty$, and $M_W$ becomes very light compared to $M_P$.
 In this limit the closed string states (almost) decouple, and the background geometry is well approximated by the $AdS_5\times S^5$ geometry.
The spin-two open string fields $w_{\mu\nu}$ become very light, i.e. almost massless, and they are very weakly coupled: $g_W\rightarrow 0$. Therefore this limit describes an (almost) conformal field theory
on the N D3-branes, with two kinds of open strings: $U(N)$ Yang-Mills gauge fields and (almost) free spin-two fields $w_{\mu\nu}$. 
Hence, all fields can be made weakly coupled, and hence this limit is well-defined and feasible.
Note that $M_W<<M_P$ can be alternatively obtained by keeping $r$ finite, but sending $g_s\rightarrow 0$. Then the string theory is weakly coupled and again $g_W\rightarrow 0$.
Small $g_s$ in fact implies that  the string scale  $M_s$ in string units is small compared to the ten-dimensional Planck mass.

\section{Holographic aspects between spin-two on the boundary and spin-four in the bulk}

%Let us notice that 
All  open string degrees of freedom/excitations on a D3-brane have a holographic description on $AdS_5$. Moreover,  
 the AdS/CFT correspondence is not only true for the massless states, but rather for the entire string modes. 
%In particular, the entire open strings on D3-branes are holographically dual to closed strings on $AdS_5$.
We will discuss in this section some aspects of the holography between the first excited open strings, namely the ${\cal N}=4$ Weyl multiplet,
  and the first excited  ${\cal N}=8$ spin-four supermultiplets of
  the  closed superstring.
  
The AdS/CFT correspondence is a duality between open strings on a d-dimensional boundary space and closed strings in a (d+1)-dimensional bulk space. The most famous
example is 4-dimensional ${\cal N}=4$ super-Yang-Mills gauge theory located on a stack of N D3-branes, which is holographically dual to ${\cal N}=8$ supergravity on
$AdS_5\times S^5$. Hence for holography to work in general, it is important to consider a limit in string theory, where all closed string modes decouple from the boundary theory.
Furthermore we need an (almost) superconformal field theory on the boundary, which possesses the same symmetries as the bulk $AdS_5$ background geometry.
More precisely, on the boundary we deal with a superconformal field theory, with superconformal symmetry group  $SU(2,2/4)\times SU({\cal N})$, where 
 $SU({ 4})$    is the R-symmetry group.
This agrees with the symmetry group of ${\cal N}=8$ supergravity on $AdS_5$.

Here we want to describe a possible way, how to include also the open string Weyl-supermultiplet $w_{\mu\nu}$ into the ${\cal N}=4$ $\leftrightarrow$ ${\cal N}=8$ boundary-bulk holography.
Limit A also not suitable for holography, since closed strings are not decoupled on the brane.
Limit A corresponds to the standard AdS/CFT correspondence, namely to the hologrographic duality between the massless spin-one gauge fields on the 4-dimensional
boundary and the massless spin-two
gravitons in the 5-dimensional bulk.
Instead we will focus on the limit B and in particular on the limit C,
where closed string  gravity on the boundary is decoupled via $M_P\rightarrow\infty$, whereas the string scale $M_s=M_W$ is kept very small compared to the
Planck mass, which means that we are considering a large  extra volume scenario in string theory.
Then the 4-dimensional, non-standard spin-two sector on the boundary possesses an (almost) superconformal symmetry and is supposed to be holographically dual to closed strings in the 
5-dimensional $AdS_5$ bulk space.

Generally 
in holography, each field $\phi(x)$ propagating on $AdS$ space is in a one to one correspondence with an operator ${\cal O}(x)$ in the field theory, which are 
coupled together by a term $\int d^4x\phi(x){\cal O}(x)$. For a rank $s$
symmetric traceless tensor, there is the following relation between the corresponding mass of the field  in the $(d+1)$-dimensional bulk and the scaling dimension $\Delta$ and the spin $s$ of the operator in the conformal field theory on the $d$-dimensional boundary:
\begin{equation}
m^2\alpha' = (\Delta+s-2)(\Delta-s+2-d)\, .\label{holomass}
\end{equation}
This formula is consistent  with the unitarity bound, which is given as 
\begin{equation}
\Delta\geq s-2+d\, . \label{bound}
\end{equation}

According to the standard holographic dictionary,
 %of \cite{Gubser:1998bc,Witten:1998qj}, 
 the most relevant operator is the conserved boundary  energy momentum tensor 
 $T_{\mu}^{\nu}$, which has conformal dimension $\Delta=4$ and spin $s=2$ and hence it saturates the unitarity bound in four dimensions.
$T_{\mu}^{\nu}$ is is coupled to a symmetric tensor $g_{\mu\nu}$,
which becomes the massless spin-two graviton field in the higher-dimensional bulk theory.\footnote{In case
the energy momentum tensor is non-conserved and has dimension 
$\Delta>2+s$, the corresponding bulk spin-two field becomes massive \cite{Bachas}.}
In our concrete case of four-dimensional super Yang-Mills theory plus Weyl$^2$ gravity given in eq.(\ref{W2}), we can derive the energy momentum tensor from the Yang-Mills action plus the 
linearized gravity action 
\begin{eqnarray}
S
=\int d^4 x\sqrt{-g}\Big{(}-{1\over 4 g_{YM}^2}  F_{\mu\nu}^aF^{a\,\mu\nu}+{2\over  g_W} 
G_{\mu\nu}w^{\mu\nu}-(w_{\mu\nu}w^{\mu\nu}-w^2) \Big{)}
. \label{W44}
\end{eqnarray}

However, let us mention that although the original (\ref{W2}) theory is invariant under conformal transformations, 
it seems that (\ref{W44}) fails as the Einstein tensor transforms non-homogeneously. Therefore, in order to restore conformal invariance of (\ref{W44}),  we have to assign a non-homogeneous transformation for the field $w_{\mu\nu}$. In fact, it can be verified that under an infinitesimal conformal transformation 
\begin{eqnarray}
 \delta g_{\mu\nu}=2\lambda(x) g_{\mu\nu},
 \end{eqnarray} 
the Einstein tensor transforms as 
\begin{eqnarray}
\delta G_{\mu\nu}=-2\nabla_\mu\nabla_\nu \lambda+2\Box \lambda g_{\mu\nu}.
\end{eqnarray}
Then, it can be verified \cite{Deser:2001pe} that the action (\ref{W44}) is invariant if $w_{\mu\nu}$ transforms as
\begin{eqnarray}
\delta w_{\mu\nu}= -\frac{2}{g_W}\nabla_\mu\nabla_\nu \lambda=\nabla_\mu \xi_\nu+\nabla_\nu \xi_\mu
\end{eqnarray}
where
\begin{eqnarray}
\xi_\mu=-\frac{1}{g_W}\nabla_\mu\lambda.
\end{eqnarray}
In other words, under a conformal transformation, the field $w_{\mu\nu}$ transforms as it would transform under a diffeomorphism generated by the gradient of the conformal factor. 

It is straightforward to calculate the energy-momentum tensor for the theory (\ref{W44}) which turns out to be 
  \begin{equation}
T_{\mu}^{\nu}= T^{\mu\nu}_F+T^{\mu\nu}_w,
\end{equation}
where 
\begin{eqnarray}
T^{\mu\nu}_F&=&\frac{1}{g_{YM}^2}\left(F^{a\,\mu}_{~~\,\rho}F^{a\,\nu\rho}-\frac{1}{4}g^{\mu\nu}F^a_{\rho\sigma}F^{a\,\rho\sigma}\right)
,\\
T^{\mu\nu}_w&=&\frac{2}{g_W}\bigg\{\Box w^{\mu\nu}-\nabla_\sigma \nabla^\nu w^{\mu\sigma}- \nabla_\sigma \nabla^\mu w^{\nu\sigma}+R^{\mu\nu} w-R w^{\mu\nu}+\nabla^\mu\nabla^\nu w+2\Big(w^{\mu\rho}w^\nu_{~\rho}- w w^{\mu\nu}\Big)\nonumber \\
&&-2\Big(G^\mu_{~\rho}\omega^{\nu\rho}+G^\nu_{~\rho}\omega^{\mu\rho}\Big)+g^{\mu\nu}\Big(G_{\rho\sigma}w^{\rho\sigma}-\frac{g_W}{2}(w_{\rho\sigma}w^{\rho\sigma}-w^2)+\nabla_\sigma \nabla_\rho w^{\rho\sigma}-\Box w\Big)\bigg\}
\end{eqnarray}
%
%
%{\bf Provide the exact expression for $T_{\mu}^{\nu}$. Also include $M_P$, $g_{YM}$ and $g_W$ in the correct way.}
The equation of motion for $w_{\mu\nu}$ is 
\begin{eqnarray}
w_{\mu\nu}=\frac{2}{g_W}S_{\mu\nu},
\end{eqnarray}
where 
$S_{\mu\nu}$ is the Schouten tensor
\begin{eqnarray}
S_{\mu\nu}=\frac{1}{2}\left(R_{\mu\nu}-\frac{1}{6}R g_{\mu\nu}\right),
\end{eqnarray} and it turns out that  $T^{\mu\nu}_F$ on-shell is 
\begin{eqnarray}
T^{\mu\nu}_F=\frac{16}{g_W} B_{\mu\nu},
\end{eqnarray}
where 
\begin{eqnarray}
B_{\mu\nu}=\nabla^{\rho}\nabla^\sigma W_{\mu\rho\nu\sigma}
+\frac{1}{2}R^{\rho\sigma}W_{\mu\rho\nu\sigma}
\end{eqnarray}
is the Bach tensor. The latter is symmetric, traceless and divergence-free
\begin{eqnarray}
B^\mu_{~\nu}=0, ~~~\nabla^\mu B_{\mu\nu}=0,
\end{eqnarray}
and therefore, $T^{\mu\nu}_F$ is also traceless (due to conformal invariance) and divergence-free (due to diff invariance). In addition $B_{\mu\nu}$ transforms under a conformal transformation 
$g_{\mu\nu}\to \Omega^2 g_{\mu\nu}$ as 
\begin{eqnarray}
B^{\mu}_{~ \nu}\to \Omega^{-4}B^{\mu}_{~ \nu}
\end{eqnarray}
and therefore it has dimension $\Delta_B=4$ (as the energy-momentu tensor).

Next we proceed to the massive spin-four operators  on the boundary in dimension $d=4$, which are coupled  to  massive spin-four, closed string fields in the bulk.
In order to be massive their scaling dimension $\Delta$ should  be larger than 6. 
These fields will become massless in the limit $\alpha'\rightarrow \infty,$ i.e. $M_s\rightarrow 0$.
In our concrete case, the relevant spin-four operator  $J^{\mu\nu\rho\sigma}$ could be for example 
\begin{eqnarray}
J^{\mu\nu\rho\sigma}&=&{\rm ST}[B^{\mu\nu}B^{\rho\sigma}]=B^{\mu\nu}B^{\rho\sigma}+B^{\rho\nu}
B^{\mu\sigma}+B^{\sigma\nu}B^{\rho\mu}\nonumber \\
&&-\frac{1}{2}\left(
g^{\mu\rho}B^{\alpha\nu} B^\rho_{\,\alpha}+
g^{\mu\sigma}B^{\alpha\nu} B^\sigma_{\,\alpha}
+g^{\nu\rho}B^{\alpha\mu} B^\rho_{\,\alpha}+g^{\nu\sigma}B^{\alpha\mu} B^\sigma_{\,\alpha}\right),
\end{eqnarray}
where ${\rm ST[]}$ denotes symmetric traceless. Other spin-four operators are
%$T^{\mu\nu\rho\sigma}$
\begin{eqnarray}
J^{\mu\nu\rho\sigma}={\rm ST}[T^{\mu\nu}_{~~\,\alpha\beta}T^{\rho\sigma\alpha\beta}],
\end{eqnarray}
or products of the Weyl tensor, as for example
\begin{eqnarray}
J^{\mu\nu\rho\sigma}={\rm ST}[ W^{\mu\alpha\gamma\kappa}
W^\rho_{~\,\alpha\delta\kappa}W^\nu_{~\, \beta\gamma\lambda}
W^{\sigma \beta\delta\lambda}],
\end{eqnarray}
where 
\begin{eqnarray}
T_{\mu\nu\rho\sigma}&=&\frac{1}{4}\left(
\tensor{W}{^\lambda_\nu_\mu^\kappa}\tensor{W}{_\lambda_\sigma_\rho_\kappa}
+\frac{1}{2}\tensor{\epsilon}{^\lambda_\nu_\tau_\xi}
\tensor{\epsilon}{_\lambda_\sigma^\chi^\psi}\tensor{W}{^\tau^\xi_\mu^\kappa}
\tensor{W}{_\chi_\psi_\rho_\kappa}
%{{W_{\nu}}^{\lambda}}_{\sigma}}^\kappa+{1\over 4}{\epsilon_{\mu\lambda}}^{\zeta\xi}{{\epsilon_{\nu}}^{\lambda\tau}}_\kappa
% W_{\zeta\xi\rho\eta}{{{W_{\tau}}^{\kappa}}_{\sigma}}^\eta
 \right)\nonumber \\
 &=& \frac{1}{4}\left(  \tensor{W}{^\lambda_\nu_\mu^\kappa}\tensor{W}{_\lambda_\sigma_\rho_\kappa}+\tensor{W}{^\lambda_\sigma_\mu^\kappa}\tensor{W}{_\lambda_\nu_\rho_\kappa}-\frac{1}{2}
 g_{\nu\sigma} \tensor{W}{^\lambda^\tau_\mu^\kappa}\tensor{W}{_\lambda_\tau_\sigma_\kappa}
 \right).
 \label{t0}
\end{eqnarray}
is  the Bel-Robinson tensor \cite{BR}. 
The  dimension of the latter is 
$\Delta_T=4$ as under conformal transformations, it transforms as   
\begin{eqnarray}
T^{\mu\nu}_{~~\rho\sigma}\to \widehat{T}^{\mu\nu}_{~~\rho\sigma}=
\Omega^{-4} T^{\mu\nu}_{~~\rho\sigma}.
\end{eqnarray}
The operators $J^{\mu\sigma\nu\rho}$ above have spin $s=4$, they transform under conformal transformations as  
\begin{equation}
J^{\mu\sigma}_{~~\,\nu\rho}\to \, 
\Omega^{-8} J^{\mu\sigma}_{~~\,\nu\rho},
\end{equation}
and their dimension is therefore $\Delta_J=8$.
 Hence
these operators  are then holographically coupled to  massive spin-four fields  in the bulk, such that  we are dealing with a higher spin-four theory in the bulk.
%Via holography,
 %$J_{\mu\nu\rho\sigma}$ is coupled to a closed string,  spin-four excitation in the bulk. 
In string theory, $J_{\mu\nu\rho\sigma}$ can be viewed as massive composite field with mass square $m^2=20/\alpha'$, corresponding to the product of two closed string graviton vertex operators. Note that this mass
is the mass on $AdS_5$, which is not the same as the mass of the corresponding string state on a flat Minkowski background.
 In the supersymmetric case,
 the field content and the 
supermultiplet structure is precisely as the one given in section 3.2.2, which is obtained by the tensor product of two ${\cal N}=4$ super-Weyl multiplets.
Since in the decoupling limit, the spin-two fields $w_{\mu\nu}$ are free fields on the 4D boundary, also the spin-four field in the $AdS_5$ bulk space should be a free field, with 
the following free field equation:
\begin{eqnarray}
&&\left(\nabla^2+\frac{3}{10}R-m^2\right)\Phi_{MNK\Lambda}=0,\nonumber \\
&&\nabla^M\Phi_{MNK\Lambda}=\Phi^{M}_{~NK\Lambda}=0, ~~~M,N,\cdots=0,1,\cdots 4, 
\end{eqnarray}
where $R$ is the scalar curvature of the $AdS_5$ space.\footnote{In general, a spin-s field in $(A)dS_d$ is described by a totally symmetric, traceless and divergentless tensor $\Phi_{M_1\cdots M_s}$ and obeys the equation \cite{Deser-1,Metsaev}
\begin{eqnarray}
\bigg[(\nabla^2+(s^2+s(d-6)+6-2d)\frac{R}{d(d-1)}-m^2\bigg]\Phi_{M_1\cdots M_s}=0,~~\nabla^{M_1}\Phi_{M_1\cdots M_s}=
\Phi^{M_1}_{~M_1\cdots M_s}=0, ~~M_i\cdots=0,1,\cdots d-1. 
\nonumber
\end{eqnarray}
. }

So in summary, the  Yang-Mills energy momentum tensor $T_{\mu\nu}$ couples to a spin-two field in the bulk, the standard graviton on $AdS_5\times S^5$, whereas $J_{\mu\nu\rho\sigma}$ couples
to a spin-four field in the bulk.
 It means in particular when 
  considering just the ${\cal N}$-extended (Weyl)$^2$ supergravity theory in four dimensions without the Yang-Mills part   that this theory   is the holographically dual
boundary theory
of an $AdS_5$ bulk theory, which is a higher spin theory with a spin-four multiplet of the $2{\cal N}$-extended supersymmetry algebra in five dimensions.
These kind of theories,  denoted by W-supergravities, were recently constructed \cite{Ferrara:2018iko} in flat four-dimensional space-time using a double copy  construction. 
Therefore, 
  the (almost) massless spin-two fields $w_{\mu\nu}$ are conjectured to be dual 
to ${\cal N}=8$ spin-four fields on $AdS_5\times S^5$. 
To support this conjecture it would be important to compute
 some correlation functions of  $J_{\mu\nu\rho\sigma}$ on the boundary and compare them with the corresponding spin-four correlation functions in the bulk.
 
 \section{Conclusions}
 
 In this paper we have discussed a special version of ${\cal N}=4$ supersymmetric  bimetric gravity coupled to ${\cal N}=4$ super Yang-Mills gauge theory.
 We have argued that, just like the open string Yang-Mills gauge fields, the massive spin-two graviton supermultiplet originates from open string excitations on D3-branes and hence
 is localized in four space-time dimensions, whereas the standard massless spin-two graviton supermultiplet is coming from the closed string sector.
 We then argued that effective action of this bimetric theory is given by the four-derivative, ${\cal N}=4$ supersymmetric Weyl$^2$ action, whose
 Weyl-supermultiplet  precisely embraces the same
 number of degrees of freedom as the first massive open string excitations on the D3-branes.
 In the massless limit, where the mass of the open string "gravitons" and their superpartners go to zero, the theory becomes ${\cal N}=4$ superconformal.
 We discussed that the holographic description of this quadratic spin-two superconformal gravity on the four-dimensional  boundary is given in terms of a higher ${\cal N}=8$
 spin-four theory in the $AdS_5$ bulk space. We have constructed the corresponding ${\cal N}=8$ spin-four supermultiplet in terms of massive closed
 string excitations in four space-time dimensions,
 which then can be lifted to the five-dimensional $AdS_5$ space. In addition
 we have identified certain spin-four operators on the four-dimensional boundary space, which, following the 
 holographic dictionary, can couple to the spin-four fields in the five-dimensional bulk.
 
 At the end of the paper, we like to close with the following additional remarks:

 \begin{itemize}

 \item
It is clear from string theory that the massive open string spin-two state cannot be a ghost state.
So eventually one has to write down an effective action for this spin-two state, which is ghost-free.
But here we are truncating the spectrum to the first excited level and neglecting all the higher open string excitations.
In the same way we are restricting the effective action to be just  with four derivatives, but we neglect all the higher derivative interactions \cite{wein1}.
It is now still a conjecture that the full effective action action of this open string spin-two state can be written as
an infinite power series expansion of the Weyl-tensor.  In fact it was recently argued in \cite{Gording:2018not}
that adding an infinite series of curvature tensors should provide an action which propagated a ghost-free open string spin-two particle.
Truncating this series to Weyl$^2$, the spin-two particle  becomes a ghost.

 \item
 In case we are dealing with a stack of N D3-branes, the massive ${\cal N}=4$ Weyl supermultiplet is colored, just like the $U(N)$ gauge fields. Therefore
 one is dealing with $N^2$ copies of interacting spin-two Weyl supermuliplets. In this paper we have considered the simpler case of just one single, neutral Weyl supermuliplet,
 which belongs to the $U(1)$ part of the $U(N)$ symmetry group, or simply is the relevant open string excitation for the case $N=1$.
 
 \item
 It would be interesting to compute the string scattering amplitudes between the massless and massive string fields using techniques already applied in  \cite{Feng:2010yx}
 in order to confirm the effective Weyl$^2$ action and the couplings between the Yang-Mills and the Weyl sectors, proposed in this paper.
 
 \item
The massive closed string spin-four field can be viewed as a kind of a bound state of two massive open string spin-two states, in analogy to the massless closed string graviton, which can
be regarded as the bound state of two massless open string gauge bosons. This observation relies in the structure of the string vertex operators and is also the basis
of the double copy constructions, which was  recently also worked out four the spin-four case \cite{Ferrara:2018iko}.
 
 \item
 It should be possible to perform  a socalled S-fold projection, getting completely
 get rid off the massless Yang-Mills sector. In this case one would entirely deal with strongly coupled,
  massive Weyl$^2$ supergravity on the boundary and with massive spin-four
 supergravity in the bulk, a theory denoted by W-supergravity, recently constructed in  \cite{Ferrara:2018iko}.
 In the massless, superconformal limit, the spin-four W-supergravity on $AdS_5$ also becomes massless.

 \item
 The scalar potential should capture also the solutions which are not the one of Einstein supergravity.
 In the bosonic case these are
the solutions where the Bach tensor vanishes but not the Ricci tensor.
While the first break conformal to Poincare supergravity,
the others may also break supersymmetry even partially, which still has to be discovered yet.
It is likely
 that any conformally flat space is a solution with vanishing Bach tensor so  it is conceivable that
AdS or even dS space are solutions of massless Weyl supergravity, as it is true in the simplest bosonic case.
 
 \item Finally, we would like to stress that the superconformal symmetry of the supersymmetric Weyl\textsuperscript{2} theory is a classical symmetry \footnote{It has been conjectured in \cite{KvP} that ${\cal N }= 4$ Poincar\'e supergravity has also a hidden superconformal symmetry.}.  Althought  such theories are power-counting renormalizable,  their one-loop beta-functions are  be non-vanishing \cite{FTs1} and therefore they suffer from a conformal anomaly. The latter leads to serious problems since conformal symmetry is gauged in Weyl gravity and therefore leads to inconsistencies \cite{Duff,FTs2,Ts3}. The same conclusion can be drawn by considering the chiral gauge anomalies of the $SU(4)$ R-symmetry \cite{RvN} and  recalling that all anomalies are accommodated  in the same multiplet of the ${\cal N} = 4$ superconformal symmetry.

 \end{itemize}

\vskip .3in

\section*{Acknowledgements:}

We like to thank Costas Bachas, Elias Kiritsis, Stefan Theisen and Timo Weigand for useful discussions.
Furthermore we gratefully knowledge enlightening discussions with Augusto Sagnotti on the Scalar Potential of Massive Weyl Supergravity and with
Angnis Schmidt-May on Bimetric Gravity.
%with N. Boulanger and F. Farakos.
 The work of S.F. is supported in part
by CERN TH Department and INFN-CSN4-GSS. The work of D.L. is supported by the ERC Advanced Grant ``Strings and Gravity" (Grant No. 320045) and the Excellence Cluster Universe.
A.K. is supported by the GSRT under the EDEIL/67108600.

\end{document}